\documentclass[sigconf,nonacm]{aamas}

\usepackage{balance} 

\usepackage{algorithm} 
\usepackage{algpseudocode}

\makeatletter
\gdef\@copyrightpermission{
  \begin{minipage}{0.2\columnwidth}
   \href{https://creativecommons.org/licenses/by/4.0/}{\includegraphics[width=0.90\textwidth]{by}}
  \end{minipage}\hfill
  \begin{minipage}{0.8\columnwidth}
   \href{https://creativecommons.org/licenses/by/4.0/}{This work is licensed under a Creative Commons Attribution International 4.0 License.}
  \end{minipage}
  \vspace{5pt}
}
\makeatother

\setcopyright{ifaamas}
\acmConference[AAMAS '25]{Proc.\@ of the 24th International Conference
on Autonomous Agents and Multiagent Systems (AAMAS 2025)}{May 19 -- 23, 2025}
{Detroit, Michigan, USA}{Y.~Vorobeychik, S.~Das, A.~Nowé  (eds.)}
\copyrightyear{2025}
\acmYear{2025}
\acmDOI{}
\acmPrice{}
\acmISBN{}

\acmSubmissionID{908}

\title[Conditional Max-Sum for Asynchronous Multiagent Decision Making]{Conditional Max-Sum for 
\\Asynchronous Multiagent Decision Making}

\titlenote{This is the extended version with the Appendix of the paper appearing in the Proc. of the 24th International Conference on Autonomous Agents and Multiagent Systems (AAMAS 2025). The accompanying codebase is available at \url{https://bitbucket.org/dtrou/conditional-max-sum-in-lane-free-traffic}.}

\author{Dimitrios Troullinos}
\affiliation{
  \institution{Technical University of Crete}
  \city{Chania}
  \country{Greece}}
\email{dtroullinos@tuc.gr}

\author{Georgios Chalkiadakis}
\affiliation{
  \institution{Technical University of Crete}
  \city{Chania}
  \country{Greece}}
\email{gchalkiadakis@tuc.gr}

\author{Ioannis Papamichail}
\affiliation{
  \institution{Technical University of Crete}
  \city{Chania}
  \country{Greece}}
\email{ipapamichail@tuc.gr}

\author{Markos Papageorgiou}
\affiliation{
  \institution{Technical University of Crete}
  \city{Chania}
  \country{Greece}}
\email{mpapageorgiou@tuc.gr}

\begin{abstract}
In this paper we present a novel approach for multiagent decision making in dynamic environments based on Factor Graphs and the Max-Sum algorithm, considering {\em asynchronous} variable reassignments and distributed message-passing among agents. Motivated by the challenging domain of lane-free traffic where automated vehicles can communicate and coordinate as agents, we propose a more realistic communication framework for Factor Graph formulations that satisfies the above-mentioned restrictions, along with {\em Conditional Max-Sum}: an extension of Max-Sum with a revised message-passing process that is better suited for asynchronous settings. The overall application in lane-free traffic can be viewed as a hybrid system where the Factor Graph formulation undertakes the {\em strategic} decision making of vehicles, that of desired lateral alignment in a coordinated manner; and acts on top of a rule-based method we devise that provides a structured representation of the lane-free environment for the factors, while also handling the underlying control of vehicles regarding core operations and safety.  Our experimental evaluation showcases the  capabilities of the proposed framework in problems with intense coordination needs when compared to a domain-specific baseline without communication, and an increased adeptness of Conditional Max-Sum with respect to the standard algorithm.
\end{abstract}



\begin{document}


\pagestyle{fancy}
\fancyhead{}


\maketitle 


\begin{figure}[t]
    \centering
    \includegraphics[scale=0.28]{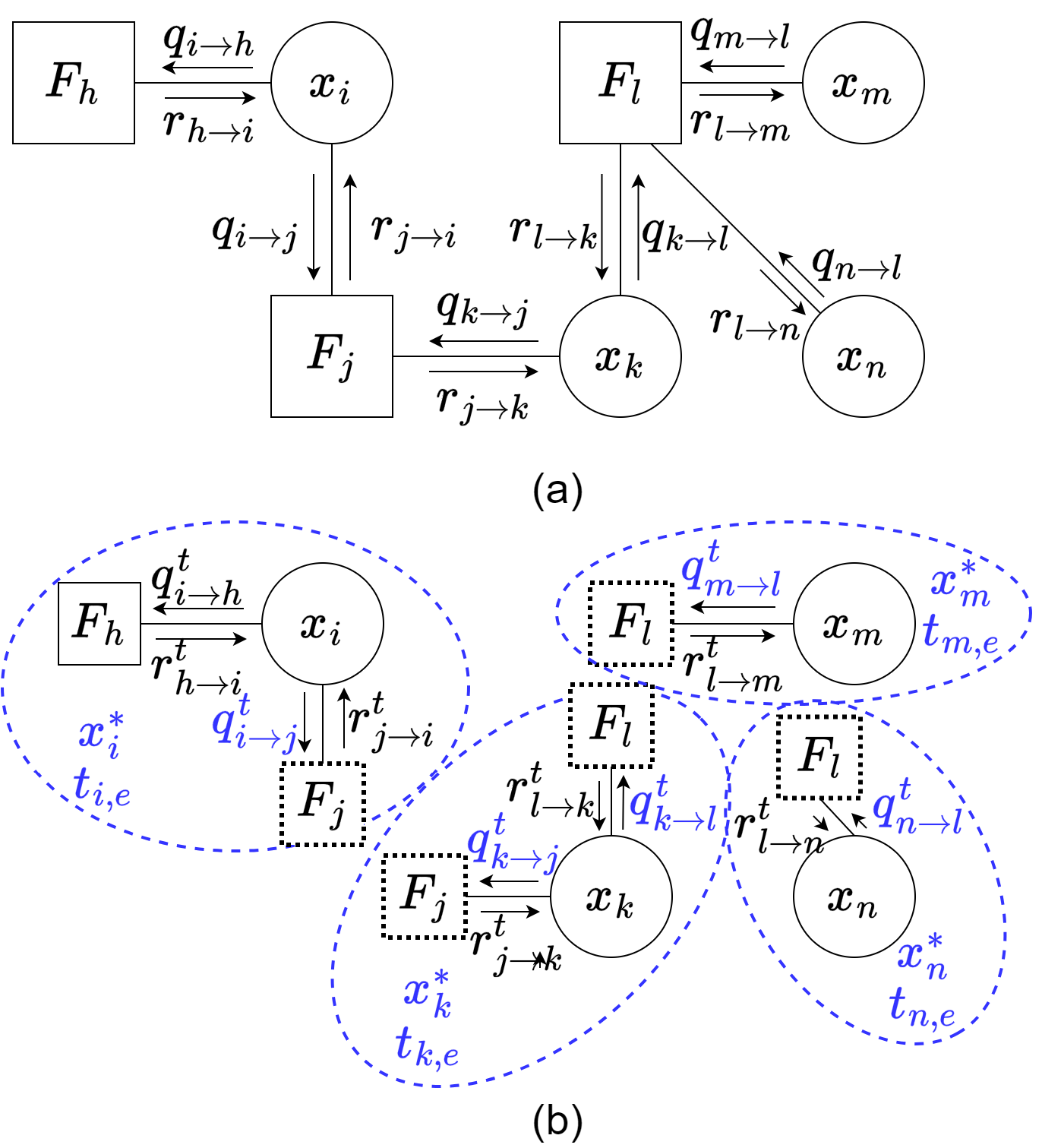}
    \caption{(a) An example of a typical FG; (b) The corresponding distributed structure for FGs.}
    \label{fig:fg_dist}
    \Description{On the top, there is an example of a typical Factor Graph containing factors, variables and illustrating the exchanged messages based on the Max-Sum algorithm, and on the bottom, we have the proposed distributed structure of the same Factor Graph.}
\end{figure}

\section{Introduction}\label{sec:intro}

Distributed Constraint Optimization Problems (DCOPs)~\cite{dcopsurvey} effectively tackle multiagent decision making in large problems that can be formulated with a highly decomposable structure.
DCOPs have been widely studied with several extensions that address broader domains such as dynamic and/or sequential environments.
Algorithmic solutions for Dynamic DCOPs usually expect that agents update their existing configuration concurrently in an online dynamic environment, or that multiple message-passing iterations can be taken 
before agents update their current configuration as a group.
These limitations can hinder the actual use of DCOPs in many multiagent environments requiring more flexible frameworks that incorporate communication or timing-related restrictions.
Existing research either tackles the asynchrony of the agents' decision making problem without taking into account large-scale and open distributed environments, by resorting to pseudo-tree graph structures that are more immutable than Factor Graphs~\cite{odpop_petcu_aaai}; or requires the synchronization of agents' decisions at every time-step~\cite{dcopsurvey}.

\begin{figure}[t]
    \centering
    \includegraphics[width=0.45\textwidth]{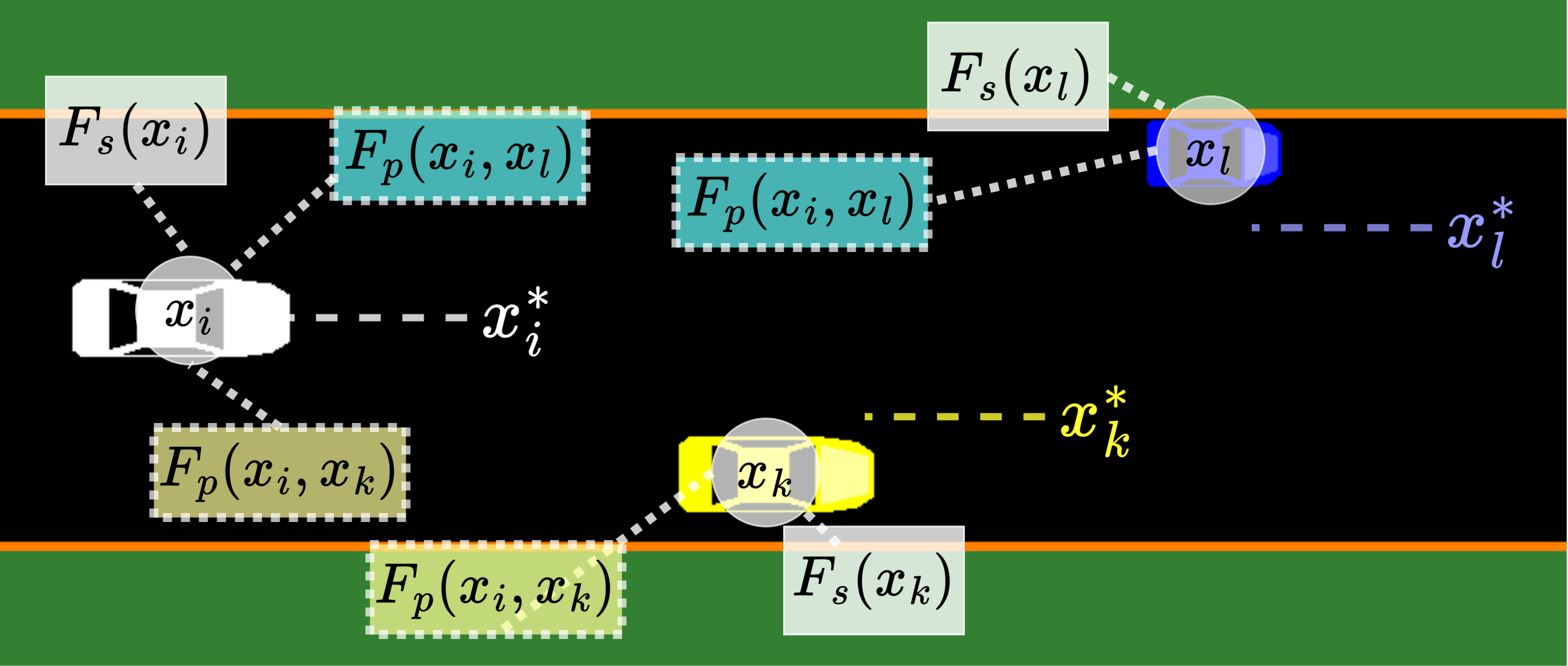}
    \caption{FG example of connected lane-free vehicles.}
    \label{fig:connected_vehs}
    \Description{Example of three lane-free vehicles on a highway, and illustration of the distributed Factor Graphs application for the coordination of their lateral alignment.}
\end{figure}

To this end, we propose a distributed communication framework for Factor Graphs (FGs) that relies on {\em information broadcasting} for communication among agents, in a way that treats agents as independent entities and not as a unified system.

A visual illustration of the framework juxtaposed with the corresponding conventional FG (contains variables $x_i$ that agents control, and decomposed factors $F_j$ of local utilities depending on variables $x_i$) can be found in Fig.~\ref{fig:fg_dist} (discussed in length in Sec.~\ref{sec:async_max_sum}).
FG formulations are often tied with the Max-Sum algorithm, which showcases great adaptability under different environments~\cite{maxsumfarinelli,Ramchurn2010,troullinos_ijcai}, and through its numerous extensions
~\cite{stranders2009,farinellibounded2011,zivan2023asynch} that solve different types of DCOPs.
In a similar vein, we put forward an extension of Max-Sum, termed as {\em Conditional Max-Sum}, 
which addresses the asynchrony on the agents' variable updates.
Conventional Max-Sum is not fully fitted for asynchronous decision-making since it is designed for static environments; while existing research on Asynchronous Max-Sum~\cite{zivan2023asynch} has a different focus, as it examines the effect of asynchronous communication schemes in static  problems, instead of asynchronous updates in sequential problems.

The proposed approach is applied in {\em lane-free traffic}~\cite{lane_free_journal}, a novel paradigm that investigates traffic environments where autonomous vehicles can fully utilize the lateral road capacity and do not obey the lane principle.
This gives rise to a challenging multiagent coordination environment, with a FG formulation that we illustrate in Fig.~\ref{fig:connected_vehs}.
There, the vehicles have  to coordinate their lateral placement through the control variable $x_i$, and update their desired lateral placement $x_i^*$, i.e., their target to reach.
Existing work~\cite{troullinos_ijcai} in lane-free driving as a multiagent problem relies on the conventional framework of FGs and Max-Sum.
As such, it does not take into account realistic restrictions regarding distributed communication and asynchrony of an open environment---especially in this setting where each agent is an independent moving vehicle with observations and dependencies that constantly change over time. 

Summing up our contributions, here we propose a distributed communication framework for FGs which allows for asynchronous decision making, giving rise to the {\em Conditional Max-Sum} algorithm. Moreover, we instantiate our framework in a realistic lane-free traffic environment for which we devise a novel formulation based on FGs.
Our experimental evaluation demonstrates the coordination efficacy of the overall framework in this domain, and autonomous vehicles under Conditional Max-Sum exhibiting an increased ability to respond quickly and better target their desired speed objective with smoother lateral maneuvers.


\section{Background and Related Work}\label{sec:back_rw}

\subsection{Factor Graphs and the Max-Sum Algorithm}
    
Factor Graphs (FGs)~\cite{kschischang2001factor} originate from probabilistic graphical models but are also well integrated within distributed AI as a common tool for DCOP formulation~\cite{maxsumfarinelli,dcopsurvey}.
Given an FG structure as the one in Fig.~\ref{fig:fg_dist}(a), we seek to obtain all control variable configurations $x_i \in \mathbf{x}$ that maximize the sum of the factors, i.e., solve the optimization problem: $\mathbf{x}^*={\arg\max}_{\mathbf{x}}\sum_j F_j(\mathbf{s}_j)$, where $\mathbf{s}_j \subseteq \mathbf{x}$.
Note that factors $F_j$ connect different variables, and may potentially associate more than two (e.g., $F_l$).
The corresponding vector $\mathbf{s}_j$ contains all variables $x_i$ connected to factor $F_j$.
For instance, in Fig.~\ref{fig:fg_dist}(a): $\mathbf{s}_l=[x_k,x_m,x_n]^T$.
In general, the factors can depend on any subset $\mathbf{s}_j \subseteq \mathbf{x}$ of the control variables.

Max-Sum~\cite{maxsumfarinelli} is an iterative, inference-based algorithm that provides an approximate solution for this optimization problem through
a message-passing operation involving two types of messages.
The first type of messages concerns values sent from variable $i$ to factor $j$:
$q_{i \rightarrow j}(x_i) = c_{ij} + \sum_{k \in M_i \setminus j} r_{k \rightarrow i}(x_i)$, where $M_i$ is the set of factor indices that variable $i$ is connected to. 
For instance, $M_k=\{j,l\}$ in Fig.~\ref{fig:fg_dist}(a). 
As such, $q_{i \rightarrow j}(x_i)$ contains an estimate for each value of $x_i$ to be sent to factor $j$.
Essentially, $q_{i \rightarrow j}(x_i)$ performs propagation of evaluations from all the other connected factors $M_i \setminus j$,
and can be viewed as the agent's current ``intents'' regarding their final configuration of variable $x_i$.
Then, $c_{ij}$ is a normalization constant that satisfies $\sum_{x_i}q_{i \rightarrow j}(x_i)=0$.
This normalization is important in cyclic graphs since we need to bound the messages' values. 
While convergence in cyclic FGs is not guaranteed, the use of $c_{ij}$ has proven to be quite effective in many studies~\cite{kok_vlassis,loopy_belief_prop,maxsumfarinelli}.

The second type of messages involves evaluations that a variable $i$ receives from a connected factor $j$:
$r_{j \rightarrow i}(x_i)=\max_{\mathbf{s}_j\setminus x_i}  \lbrack F_j(\mathbf{s}_j) + \sum_{k \in N_j \setminus i} q_{k \rightarrow j}(x_k)  \rbrack$, where $N_j$ is the set containing the variable indices connected to factor $j$, and the maximization process involves all the variables $\mathbf{s}_j$ connected to factor $F_j$ without $x_i$.
For instance, in Fig.~\ref{fig:fg_dist}(a), $N_l=\{k,m,n\}$, meaning that variables $x_k,x_m,x_n$ are connected to factor $F_l$.
These messages calculate for each possible value $x_i$ an estimate for the outcome considering both the associated factor, and all other messages sent to it. 
Since the $q_{k\rightarrow j}$ values provide these connected $N_l$ variables evaluation for their respective decisions, agent $i$ maximizes over all these variables, in order to have an evaluation that incorporates both the immediate factor's value and other agents' influence.
In every new iteration, the previous evaluations of the messages are utilized.
This operation is performed until some stopping criterion (e.g., time,  iterations number), or if all message values converge (within a threshold).
Finally, each agent computes the optimal value as: $x_i^* = \arg \max_{x_i} \sum_{j \in M_i} r_{j \rightarrow i}(x_i)$, i.e., the $x_i$ that maximizes the sum of all the received messages.
Without loss of generality, we consider that each agent controls one variable, 
and refer (same indexing) to agents and variables interchangeably.

\subsection{Related Work}

Dynamic DCOPs (D-DCOPs)~\cite{dcopsurvey} extend conventional DCOPs to dynamic environments. 
Consequently, the notion of time is introduced to the problem. In an FG problem formulation, we consider a dynamic graph structure with the possibility to change in time. 
At a time-step $t$, the FG contains a set of factors $M$, with $F_j(\mathbf{s}_j) \in M$, and $x_i \in N$, where $N$ is the set of variables.
There are two types of changes for every update from $t$ to $t+1$: (a) existing factors $F_j(\mathbf{s}_j)$ can now contain different values; or (b) new variables $x_i$ and factors $F_j(\mathbf{s}_j)$ can be introduced, and existing ones can be removed.
Commonly, D-DCOPs do not model how the factor evolves over time and algorithmic extensions~\cite{dcopsurvey} re-solve every new static instance of the problem at time $t$, effectively incorporating domain knowledge to improve performance~\cite{Ramchurn2010}.
Prior work in lane-free environments~\cite{troullinos_ijcai} (and~\cite{stranders2009}) also solves every new instance of the problem by relying on the previous solution as a starting point, with the assumption that changes in the graph modelling vehicles' interactions will not be substantially different.
In contrast,~\cite{Nguyen_Yeoh_Lau_Zilberstein_Zhang_2014} explicitly model the FG's evolution as Markovian D-DCOPs, thereby they tackle the sequentiality of the problem and incorporate methods from RL.
Moreover,~\cite{scalablemcts} addresses D-DCOPs from the perspective of DecMDPs, which can be viewed as a similar problem, and perform planning based on Monte-Carlo tree search with Max-Sum.\footnote{The Max-Plus algorithm (employed in~\cite{scalablemcts}) is effectively the same method with Max-Sum when the FG follows a structure that contains factors up to $2$ variables.}

Finally, we note that papers applying Max-Sum on Mobile Sensor Teams (MSTs), dealing with exploration issues~\cite{yedidson2018} or collision avoidance~\cite{pertzovsky2024collision}, exhibit conceptual similarities to our domain (which is natural since MSTs involve a dynamic environment of moving agents).
Regardless, they do not deal with asynchrony in the agents' variable updates as we do in this paper.


\section{Asynchronous Decision Making with the Max-Sum Algorithm}\label{sec:async_max_sum}

We now present a novel extension of Max-Sum for problems that can be formulated as D-DCOPs with specific constraints that render them more realistic in a large-scale distributed coordination environment.
The first restriction we impose is we cannot have multiple iterations of the algorithm, since in an open distributed environment we cannot easily assume that agents will be able to communicate multiple times before updating their decision and iteratively propagate messages.
We rather rely on a more realistic notion of {\em information broadcasting} from the perspective of each agent.
This is visualized in Fig.~\ref{fig:fg_dist}(b), where each agent broadcasts all $q$ messages to be sent to connected factors $j$ that involve other agents as well, 
with additional information (indicated with blue color) relevant to 
the Max-Sum extension we later establish.
At time-step $t$, agents first observe the broadcasted $q$ messages of nearby agents, then calculate their $r$ messages followed by the updated $q$ messages to be broadcasted.
As such, at time-step $t$, $r^t$ messages now rely on the previously broadcasted messages $q^{t-1}$.
Otherwise, we would need to either establish some form of order among agents' message updates (not realistic and potentially restrictive in large environments), or examine the effect of asynchronous message-passing  (essentially a different problem) in this setting, as put forward and studied in~\cite{zivan2023asynch}.
Therefore, each agent $i$ at time-step $t$ can update its received messages as:

\begin{eqnarray}\label{eq:r_ij_t}
      r^t_{j \rightarrow i}(x_i) = \max_{\mathbf{s}_j\setminus x_i} \Big  \lbrack F^t_j(\mathbf{s}_j) + \sum_{k \in N_j \setminus i} q^{t-1}_{k \rightarrow j}(x_k) \Big \rbrack
\end{eqnarray}

\noindent and then its $q^{t}$ messages to be broadcasted to other agents: 

\begin{eqnarray}\label{eq:q_ij_t}
    q^t_{i \rightarrow j}(x_i) &=& c_{ij} + \sum_{k \in M_i \setminus j} r^t_{k \rightarrow i}(x_i)
\end{eqnarray}

Naturally, this choice comes at the cost that a single iteration of the algorithm will probably result in a significantly subpar solution quality.
However, in realistic dynamic environments such as the lane-free traffic domain, it is not reasonable for agents to update their configuration at every time-step, since they would not be able to commit to their decision and probably result in oscillatory behaviour. 
In~\cite{petcu2007_rsdpop}, authors are motivated by the same assumption and establish the notion of commitment deadlines for agents regarding their update.
Likewise, before introducing asynchronous operation, we can provisionally define a common time-period $T$ for agents' updates, where all agents can adjust their configuration $x_i^*$ based on the received messages $r$.
More specifically, agents update $x_i^*=\arg\max_{x_i}\sum_{j \in N_i} r^t_{j\rightarrow i}(x_i)$ periodically, while in intermediate steps they can only communicate and perform one iteration of the algorithm, i.e., update with Equations~\ref{eq:r_ij_t} \& ~\ref{eq:q_ij_t}, and maintain the last update of $x^*_i$.
This effectively allows us to:{
\em (1)} perform multiple iterations before agents update their decision but now in a distributed setting, and {\em (2)} avoid indecisive behaviour of agents due to the dynamic nature of the problem.
Certainly, this assumes (as in related work) that the FG does not change abruptly within this time-period $T$ required for the update.
Otherwise, the previously exchanged messages have diminishing value.

Even with this common time-period $T$, there is a need for a {\em synchronized} clock among agents, imposing to them whenever they should update their decisions.
We additionally lift this requirement, and allow for {\em asynchronous} variable updates $x_i$ for each agent. 
This introduces a sense of autonomy as well since agents are not bound to make decisions alongside others, and can react in a timely manner depending on their local observations.
As a result, at every time-step $t$, a subset of agents update their existing configuration.
In this more flexible framework, one could simply apply the standard Max-Sum algorithm with the use of message passing operations in Equations~\ref{eq:r_ij_t} \& \ref{eq:q_ij_t}, but with the distinction that each agent updates $x_i^*$ at a potentially different time-step from others.
However, the calculation of incoming $r^t$ messages in Max-Sum (Eq.~\ref{eq:r_ij_t}) entails a  critical assumption: the local maximization from the perspective of an agent $i$ is performed based the propagated $q$ messages that reflect the final configuration, i.e., agent $i$ updating its variable at time $t$ {\em assumes} that all other agents connected to it through a factor will do so concurrently.

\subsection{Conditional Max-Sum Algorithm}\label{subsec:cond_ms}

In order to efficiently tackle asynchronous decision making in this new context, we need to redefine the message passing equation for $r^t$ messages.
For this, we request supplementary information from agents, namely a {\em time-estimate} $t_{i,e}$ from each agent $i$ regarding its next update and its last variable configuration $x^*_i$ (cf Fig.~\ref{fig:fg_dist}(b)).
As mentioned, the issue lies with the local maximization of connected variables in $r^t$  messages, due to the underlying assumption of an one-time variable configuration for all agents.
We extend this notion by performing a {\em conditional} maximization depending on the {\em relative} time-estimates between agent's $t_{i,e}$ and the other agents' estimates: $\mathbf{t}_{j,e} = \{t_{k,e} : \forall k \in N_j \setminus i\}$, and their existing configuration $\mathbf{x}_j^* = \{x_{k}^* : \forall k \in N_j \setminus i\}$.
In this manner, instead of maximizing over all connected variables $\mathbf{s}_j\setminus x_i $ on factor $F_j$, we examine for each variable  $\forall x_k \in \{\mathbf{s}_j\setminus x_i\}$ whether to: (a) include variable $x_k$ in the maximization operation; or instead (b) use directly the existing value  ${x}_k^*$ based on the broadcasted information.

This choice depends on the relative time-estimate of agent $i$ with respect to its connected agents (through a factor $j$). 
We first consider factors $F_{j}(x_i,x_k)$ connecting exactly two agents $i,k$.
At time-step $t$, both agents have broadcasted their time-estimates $t_{i,e},t_{k,e}$ and current variable assignments $x^*_i, x^*_k$.
From the perspective of agent $i$, its $r^t_{j \rightarrow i}$ messages will be updated as:

\begin{align}\label{eq:cms_r_t}
    r^t_{j \rightarrow i} (x_i) = 
    \begin{cases}
        {\max}_{x_k} \Big  \lbrack \;  F^t_j(x_i,x_k) +q^{t-1}_{k \rightarrow j}(x_k)        
        \!\Big \rbrack,        
        & \text{if $t_{k,e}-t_{i,e}\leq t_{e}$}\\        
          F^t_j(x_i,x^*_k) + q^{t-1}_{k \rightarrow j}(x^*_k), & \text{otherwise}
    \end{cases}
\end{align}

\noindent where $t_{e}$ is a positive constant accounting for the reaction time of the underlying D-DCOP when agents update their assignments.\footnote{For instance, the $x_i$ variables control the lateral placement of lane-free vehicles. A small time difference is negligible when accounting for the reaction time of the underlying system.}
To put it simply, if agent $i$ plans to update its assignment $x^*_i$ before $k$ does, then the implicit assumption on $k$'s update through the $\max_{x_k}$ operator will be less accurate than directly embedding the broadcasted assignment $x^*_k$ for the calculation.
The following example containing two agents connected with a factor illustrates this aspect and the overall reasoning for the approach.

\begin{figure}[t]
    \centering
    \includegraphics[scale=0.25]{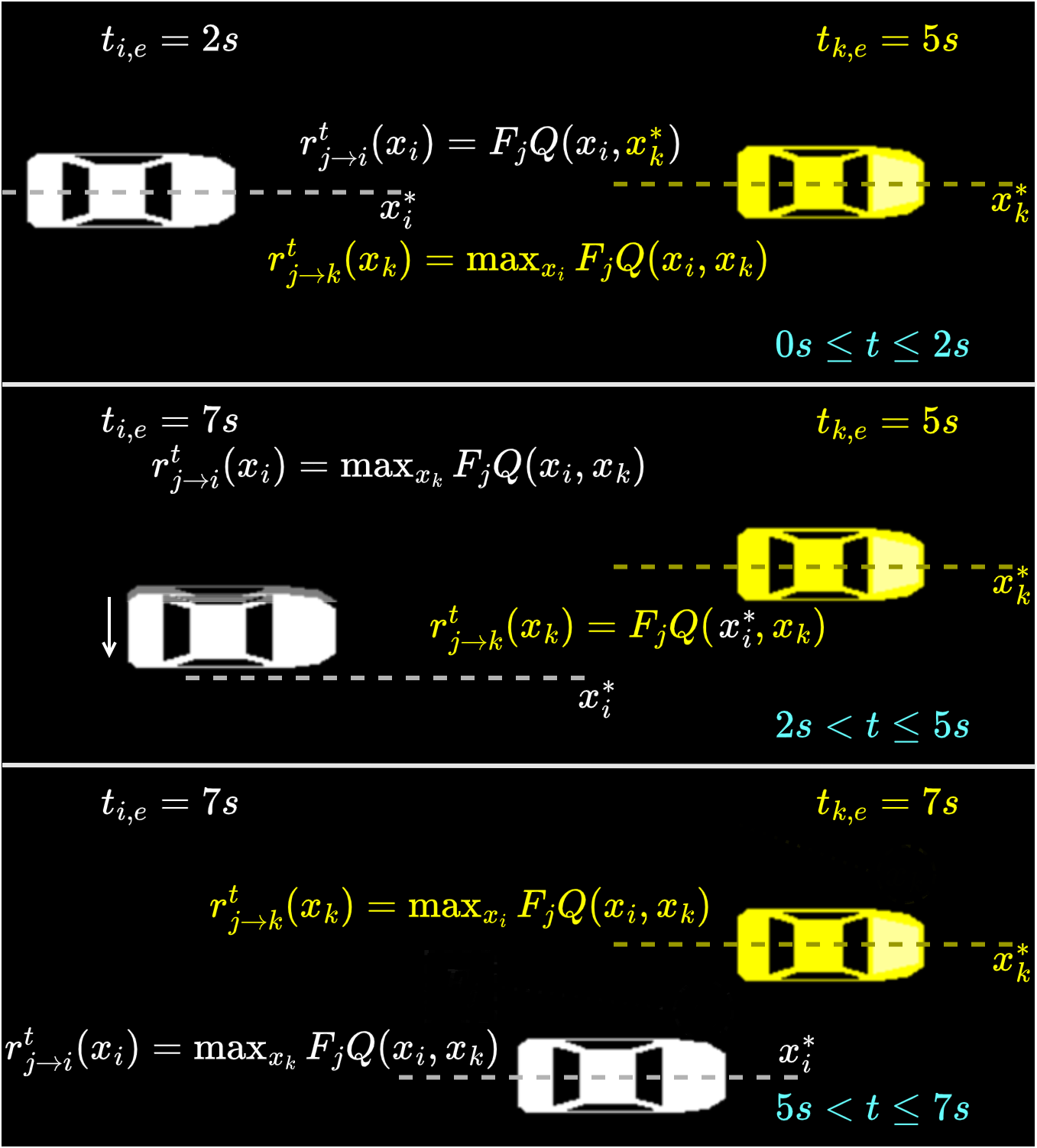}
    \caption{Illustrative example of two agents coordinating with Conditional Max-Sum to update their decisions asynchronously.}
    \label{fig:cond_ms_example}
    \Description{Illustrative example of two agents (lane-free vehicles) for the use-case of Conditional Max-Sum. The example is divided in three time phases from top to bottom, showing the gradual change in the way agents adjust the calculation of receiving messages depending on the timings of their upcoming updates.}
\end{figure}

\begin{example}[Two agents in lane-free traffic connected with a factor]\label{ex:cond_lf_overtake_example}
Consider two lane-free agents $i,k$ connected with a factor $F_j(x_i,x_k)$ at time $t=0s$. The variables control the vehicles' lateral alignment, and the factor $F_j$ aims to coordinate them so that the vehicle on the back can overtake if it desires to (see more details in Sec.~\ref{sec:mas_lf}). 
In this example, agent $i$ wishes to overtake and factor $F_j$ incorporates this information.
At time $t=0s$, the last variable assignments $x_{i}^*,x_{k}^*$ and shared time estimates $t_{i,e},t_{k,e}$ are shown in Fig.~\ref{fig:cond_ms_example} at the top segment, and the dashed lines crossing the center of each vehicle showcase where the desired lateral positioning is according to $x_{i}^*,x_{k}^*$ respectively.
For the sake of simplicity, we set $t_e=0$ in this example and consider that the time-estimates shared by agents will be fully accurate.
Additionally, for reasons of compactness in the figure, we define ${FQ}_j(x_i,x_k) = F_j(x_i,x_k) + q_{k\rightarrow j}(x_k)$.

Focusing at the top segment, agent $i$ has an earlier time-estimate for its update, meaning that it should not maximize over $x_k$ for its $r^t_{j\rightarrow i}(x_i)$, since at that time, agent $k$ will remain in its current lateral position. 
Contrariwise, $k$ plans to update its variable much later, therefore it is more rational from its end to maximize over variable $x_i$ in order to reach a decision based on $i$'s potential movement, and not its current lateral positioning according to $x_i^*$.
Then, at the middle segment the situation is inverted due to the updated time-estimates, and finally at the bottom part, both agents perform the conventional maximization of Max-Sum due to the synchrony of their upcoming reassignment.

\end{example}

As of now, we have only addressed factors connecting $2$ agents.
Notably, factors with only $1$ agent are a special case and do not require this procedure.
However, the same notion can be directly applied in larger factors as well, albeit with a compact form of Eq.~\ref{eq:cms_r_t}.
For this, we revise Eq.~\ref{eq:cms_r_t} in a way that combines the two cases as: $r^t_{j \rightarrow i} (x_i) = \max_{s_j^{t,e}}  \lbrack F^t_j(x_i,x_k) + q^{t-1}_{k \rightarrow j}(x_k) \rbrack$, where $s_j^{t,e} = \{x_k : t_{k,e} - t_{i,e} \leq t_{e}\}$. Following this, we can directly generalize for factors of various sizes accordingly:

\begin{eqnarray}\label{eq:cms_r_t_p}
    r^t_{j \rightarrow i} (x_i) = \max_{\mathbf{s}_j^{t,e}} \Big  \lbrack F^t_j(\mathbf{s}_j) + \sum_{x_k \in N_j \setminus i} q^{t-1}_{j \rightarrow k}(x_k) \Big \rbrack
\end{eqnarray}

\noindent where $\mathbf{s}_j^{t,e}$ contains only the variables in factor $j$ that should be maximized based on the relative time-estimate of agent $i$:

\begin{eqnarray}
    \mathbf{s}_j^{t,e} = \{x_k : \forall k \in N_j \setminus i, t_{k,e} - t_{i,e} \leq t_{e}\}
\end{eqnarray}

\noindent and the input vector $\mathbf{s}_j$ is formed by $\mathbf{s}_j= \{x_i \} \cup \mathbf{s}_j^{t,e} \cup \mathbf{s}_{j^*}^{t-,e}$, where: 

\begin{eqnarray}
    \mathbf{s}_{j^*}^{t-,e} =  \{x^*_k : \forall k \in N_j \setminus i, t_{k,e} - t_{i,e} > t_{e}\}
\end{eqnarray}

\noindent contains the broadcasted $x^*_k$ assignments for the connected variables that do not comply with the time-estimate criterion, i.e., for all excluded neighboring variables from $s_j^{t,e}$, the broadcasted information $\mathbf{x}^*_j$ is utilized to fill the arguments in $F_j^t(\mathbf{s}_j)$ and $q^{t-1}_{k \rightarrow j}(x_k), \forall x_k \in N_j \setminus i$ instead of maximizing over these variables as well.
In Appendix Sec.~A, we discuss the impact of this revised equation and its connections to local search-based methods.
The full algorithmic process for the distributed update of each agent per time-step is outlined in Algorithm~\ref{alg:ag_update}.
The steps concerning lines 2,5,6 of the algorithm can depend on additional domain-specific mechanisms that we later establish (see Sec.~\ref{subsec:fg_lf}) for lane-free traffic.

\begin{algorithm}[t]
    \caption{Agent $i$ Distributed Update}
    \label{alg:ag_update}
    \begin{flushleft}
      \textbf{Input}:\,Surrounding agents and their previously broadcasted information 
      $\forall j \in M_i \langle q^{t-1}_{\cdot \rightarrow j} , \mathbf{x}_j^* , \mathbf{t}_{j,e} \rangle$\\
      \textbf{Output}:\,Updated broadcasted information $\forall j \in M_i \langle {q}^{t}_{i \rightarrow j}\rangle, \langle x_i^* , t_{i,e} \rangle$
    \end{flushleft}
    \begin{algorithmic}[1]
        \State Observe surroundings and broadcasted information
        \State Update connected factors' information (for existing factors: change values/remove, or form new connections)
        \State Update $r^t$ messages from broadcasted $q^{t-1}$ messages (Eq.~\ref{eq:cms_r_t_p} for Conditional Max-Sum)    
        \State Update $q^t$ messages (Eq.~\ref{eq:q_ij_t})
        \State Decide whether to update $x_i^*$
        \State Update time-estimate information $t_{i,e}$
        \State Broadcast $q^t$ messages for all connected factors, variable assignment $x_i^*$ and time-estimate for next variable update $t_{i,e}$ 
    \end{algorithmic}
\end{algorithm}


\section{Multiagent Coordination in Lane-Free Traffic}\label{sec:mas_lf}

In this section, we present the FG formulation based on lateral regions that coordinates the vehicles' lateral alignment in lane-free traffic.

\subsection{Problem Description}\label{subsec:lf_problem}

In the examined lane-free traffic environment, each vehicle populating the road operates on a 2-dimensional space consisting of a longitudinal (front/back) and lateral (left/right) axis.
The investigated scenarios consist of a lane-free highway that is either static (contains a specific set of vehicles) or an open environment (new vehicles constantly enter the highway). 
Each vehicle possesses a separate {\em desired speed} $v_i^d$ objective that pursues, consequently resulting in many instances where vehicles wish to overtake while being surrounded by nearby traffic in a lane-free setting. 
The underlying policy of the vehicles is a rule-based method that automatically adjusts their acceleration in response to surrounding traffic. 
As such, the vehicles have by default a ``reactive'' behaviour that follows safety rules in order to avoid collisions, and do not take strategic initiatives, i.e., perform overtake maneuvers or give priority to other vehicles.
This type of behaviour is handled by our D-DCOP formulation of the problem, where the vehicles need to coordinate their lateral placement $y_i$ in order to perform cooperative maneuvers that benefit their own and/or nearby traffic's objectives.
To this end, each vehicle targets a {\em desired lateral placement} $y_i^d$ through our approach, and they are modelled as agents in a FG structure that evolves over time due to the dynamic nature of the problem; following the distributed communication framework we propose with asynchronous updates of their desired lateral placements for responding timely to their respective local situation.

\subsection{Lane-Free Traffic Environments with Dynamic Lateral Regions}\label{subsec:lat_regs}

The primary tool that enables coordination in a manner suitable for strategic coordination in lane-free traffic is that of {\em dynamic lateral regions}.
With this the vehicles can interpret the observed and/or communicated information from nearby traffic in a way that provides a {\em real-time structured representation} of the environment, and allows them to decide upon their low-level control.

\begin{figure}[t]
    \centering
    \includegraphics[width=0.48\textwidth]{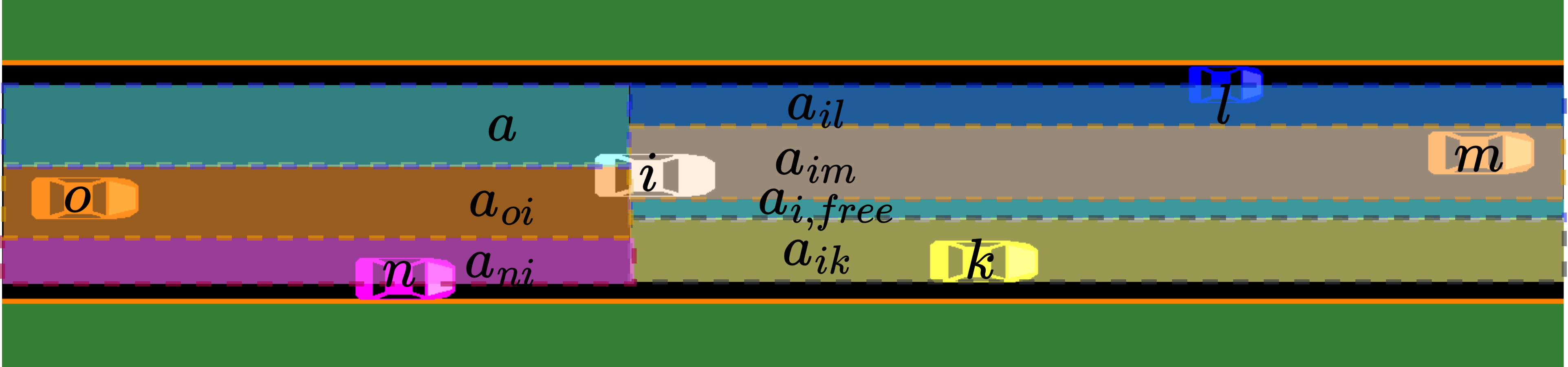}
    \caption{Formed lateral regions with acceleration estimates from the perspective of agent $i$.}
    \label{fig:lat_regions_viz}
    \Description{Illustration of the formation of lateral regions based on surrounding traffic from the perspective of an agent located at the center of the figure.}
\end{figure}

From the perspective of an ego vehicle $i$, we distinguish between upstream (vehicles on the back of $i$) and downstream traffic (vehicles on the front of $i$).
Given an observational distance, all vehicles are monitored and the lateral space is accordingly partitioned into lateral regions, as visualized in Fig.~\ref{fig:lat_regions_viz}, which correspond to where the center point of vehicle $i$ can be positioned laterally.
At any time, $i$ is located at a specific lateral region, in which its longitudinal behaviour (gas/brake) is being influenced by the front vehicle occupying this region. 
This is decided according to a car-following method as done typically in lane-based environments, with the vehicle in front as the leader to follow.
For this task, we employ the Enhanced Intelligent Driver Model (EIDM)~\cite{kesting2010enhanced}, which is an extension of one of the most popular car-following methods, that calculates the longitudinal acceleration $a_i$ of the vehicle, taking into account $i$'s desired speed $v_i^d$ while respecting a time-gap value with the vehicle in front to avoid critical situations.
In this manner, given two vehicles $(i,j)$ with $j$ being in front of $i$, we can calculate the acceleration $a_{i, j}$ of $i$ when $j$ is in front according to EIDM.
Likewise, we can calculate this acceleration for each lateral region, meaning we can have an {\em estimate} of the acceleration of ego vehicle depending on its lateral placement.
These acceleration estimates are instrumental in our approach, as they quantify the {\em value} of residing at a lateral region, and consequently the {\em benefit} of shifting laterally to a different region by simply comparing the corresponding acceleration evaluations.
Vehicles are also influenced by upstream traffic, resulting in nudging behaviour which is an important characteristic of lane-free traffic~\cite{lane_free_journal}.
The acceleration estimates of all lateral regions for downstream and upstream traffic are integrated in underlying safety rules that regulate the vehicles' control input and therefore their behaviour. A more detailed presentation can be found in  Appendix~B.

\subsection{Factor Graphs in Lane-Free Traffic}\label{subsec:fg_lf}

We can form a FG of connected vehicles as visualized in Fig.~\ref{fig:connected_vehs}, assuming the necessary communication capabilities for vehicle within close proximity.
The control variable $x_i$ for each vehicle $i$ is the lateral deviation $dy_i$ which determines the updated {\em desired lateral alignment} $y_i^d$ of the vehicle.
For all calculations relevant to the FG formulation, we examine candidate lateral positions $y_i'$ according to a value for $x_i$ accordingly: $y_i' = y_i + x_i$.
Each vehicle can be connected to two types of factors in our formulation.
First, the single factor involves only one vehicle and accounts for motivating the vehicle to remain within the road boundaries.
Its form is: $F_s(x_i) = -B_c\cdot outOfBounds_i(x_i)$, 
where the $outOfBounds_i(x_i)$ element yields a negative utility according to the coefficient $B_c$ if the examined configuration $x_i$ results in a lateral placement $y_i'$ that would exceed the road boundaries.
More importantly, the FG formulation contains a second type of a pairwise factor $F_{p}(x_i,x_j)$ that connects two vehicles $i$ and $j$, with $j$ preceding $i$. 
Its presence serves to motivate both $i$ and $j$ at moving laterally according to $i$'s {\em desire} to overtake through {\em regret minimization}.
Notably, since the factor affects both $i$ and $j$'s decision due to their involvement, they can accordingly control their lateral behaviour in a coordinated manner.
This is accomplished by the following formulation:

\begin{eqnarray}\label{eq:f_ij}
    F_{p}(x_i,x_j) = regret(x_i,x_j) + comfort(x_i,x_j)
\end{eqnarray}

\noindent where the first term is the calculated {\em regret} from the perspective of the receding agent $i$, and has the following form:

\begin{eqnarray}\label{eq:regret}
    regret(x_i,x_j) = -R_c\cdot (a_{i,free} - a_{i, j})^2 \cdot overlap_{ij}(x_i,x_j)
\end{eqnarray}

\noindent and the value $a_{i,free}$ is the calculated acceleration of $i$ when located in lateral regions without a leader, meaning that this value only accounts for the desired speed objective of the agent and that $a_{i,free}\geq a_{i,j}$.
As such, the difference $(a_{i,free} - a_{i, j})$ expresses the {\em regret} of agent $i$ for having $j$ {\em in front} of it,\footnote{$i$ observing $j$ as leader for the examined configuration $y_i',y_j'$.} using a positive coefficient $R_c$ and a negative sign to comply with the algorithm's maximization criterion.
Note that whenever this type of factor connects two vehicles, this regret value is assigned to it only for configurations of $x_i,x_j$ that would result in the agents having their lateral alignments ``overlap'' at any point during lateral deviation from the current placement $y_i,y_j$ towards the examined one $y_i',y_j'$.
More details can be found in Appendix~C.1.

The second term of Eq.~\ref{eq:f_ij} serves to mitigate unnecessary lateral deviations through a {\em comfort} utility with the form: 
$comfort(x_i,x_j) = -C_c \cdot \Bigl (|x_i| +|x_j| \Bigr )$   
with $C_c$ as a coefficient.
The tuning of $R_c,C_c$ regulates the behaviour of agents regarding overtaking and comfortable driving.
A similar idea for comfort can be found on prior work in Proactive DCOPs~\cite{hoang2022proactive} and RS-DPOP~\cite{petcu2007_rsdpop}, where authors contain an additional cost term that penalizes abrupt and unnecessary changes in variables' values.
We instead consider this as a domain-specific aspect of our approach, and do not embed it within the algorithm.

Each agent is associated with one single factor, as shown in Fig.~\ref{fig:connected_vehs}.
A pairwise factor $F_{p}(x_i,x_j)$ between agents $i,j$ is considered according to their proximity.
Additionally, each agent $i$ prunes its own connections with others in order to conform to upper limits regarding the number of pairwise connections it can have downstream and upstream, respectively.  
This process is part of line 2 in Alg.~\ref{alg:ag_update}, with more details in Appendix~C.2.

\subsection{Asynchronous Decision Updates in Lane-Free Traffic}\label{sec:async_max_sum_lf}

With the FG formulation above, we have the necessary components to model the lane-free traffic environment as a DCOP. 
However, additional elements need to be prescribed for asynchronous decision-making as discussed in Sec.~\ref{subsec:cond_ms}.    
Agents update the values of all connected factors according to their real-time observation of nearby vehicles and the information from the constructed lateral regions, but they also take into account the broadcasted assignments $\mathbf{x}_j^*$ of all factors $j$ connecting them to other vehicles.
The information for the lateral placement of vehicles is informed by these assignments so that the FG formulation integrates these in the lateral positioning $y_i,y_j$ for the calculation of the factor's values.
This is an important nuance of the formulation, as the vehicles (through the DCOP formulation) can proactively ``argue'' regarding their lateral alignment decisions and not take into account intermediate states while they maneuver from one lateral alignment to the next.

At each time-step, all agents independently decide (line 5 in Alg.~\ref{alg:ag_update}) whether to update their variable assignment $x_i^*$ based on independent time-windows $\lbrack T_{min}, T_{max} \rbrack$.
Once $T_{min}$ time has passed since the agent's last reassignment, only then $i$ examines whether it has reached the selected lateral positioning $y_i^d$ within a small error $y_e$.
If so, then it updates $x_i^*$ based on the received messages $r^t$. 
This update can also take place without the aforementioned condition being satisfied if the time-window is exhausted, i.e., $T_{max}$ time has passed since the agent's previous update.
Finally, each agent needs to provide a time-estimate for the next update in order to comply with the proposed communication scheme (line 6 in Alg.~\ref{alg:ag_update}). 
This is done in a very straightforward manner by computing the remaining time-steps to reach the selected lateral positioning $y_i^d$.
The lateral maneuver of the vehicle solely depends on the use of the movement dynamics and is deterministic.
Thereby, the future lateral trajectory of the vehicle can be fully predicted (see Appendix~C.3 for more details).
Of course, the intended lateral goal of the agent can be compromised in practice by other agents blocking its path towards it, due to safety rules for lateral alignment, as mentioned in Sec.~\ref{subsec:lat_regs}.
Consequently, the shared time-estimates cannot be fully accurate in a realistic environment under uncertainty, hence the reason they are updated at every time-step.


\section{Experimental Evaluation}\label{sec:exp_eval}

We empirically evaluate our approach by comparing $3$ different Max-Sum variants in the proposed distributed framework with asynchronous variable reassignments, along with a baseline heuristic method without communication among agents.
Specifically, we have: (a) \textbf{Max-Sum}: the standard Max-Sum algorithm; (b) \textbf{No-Max-Sum}\footnote{Reflects simpler local search-based methods, see Appendix~A.}: we rely solely on the broadcasted assignments from neighbors instead of maximizing in Eq.~\ref{eq:cms_r_t_p}, i.e., $\mathbf{s}_j^{t,e}$ is empty; (c) \textbf{Cond-Max-Sum}: Conditional Max-Sum, as in Sec.~\ref{sec:async_max_sum}; and finally (d)  \textbf{MOBIL} (baseline): Rule-based method based on the popular lane-change model MOBIL~\cite{kesting2007mobil} that does not employ any communication among agents (more information in Appendix~D.1).

Evaluation metrics for experiments contain: (a) the \textbf{average speed} of all vehicles throughout the simulation: $v_x^{avg}$ in meters per second $(m / s)$. \footnote{Note that for the examined range $\lbrack 25, 35 \rbrack m / s = \lbrack 90, 126 \rbrack km / h$.}
This metric indicates the efficiency of the vehicles' movement, with higher values suggesting more efficient behaviour, given that the desired speed goals of vehicles are not lower on average than the speed measurements.
Additionally, an integral metric we include is:
(b) the \textbf{average speed deviation} of all vehicles $v^{avg}_{dev}$ in meters per second $m/s$.
This metric effectively measures the vehicles' deviation $|v_i^x - v_i^d|$ of their current speed $v_i^x$ from their desired speed $v_i^d$ objective, and consequently how close vehicles are towards their respective desired speed goal.
Moreover, we measure: (c) the \textbf{average {\em jerk}} of vehicles in $m / s^3$ to be minimized, which is the derivative of acceleration $m / s^2$, and a commonly used metric for discomfort of passengers~\cite{huang2004jerksae}. 
As we focus on the way vehicles move laterally in the lane-free environment, we show the jerk regarding lateral maneuvers $j_y^{avg}$.
Finally, a system-level measurement is included for Sec.~\ref{subsec:exp_large}: (d) the \textbf{{total-time-spent}} ($TTS$) hours ($h$), a standard metric in transportation that depicts the accumulated travel time summed over all vehicles in simulation.\footnote{Videos demonstrating the performance of all methods can be found at: \url{https://bit.ly/4k6FXx1}.}

\subsection{Lane-Free Coordination Problem}\label{subsec:exp_small}

\begin{table}[t]
  
  \caption{Results for the coordination problem and the two flow configurations on the $2km$ open highway.}
  \label{tab:res_large_medium}  
  \scalebox{0.92}{
  \begin{tabular}{r|ccccc}
  \toprule
  Coord. Problem & \textit{$v^{avg}_x$ ($m/s$)} &\textit{$v^{avg}_{dev}$ ($m/s$)}  & \textit{$j^{avg}_y$ ($m/s^3$)} & $TTS (h)$ \\ 
    \midrule
    Max-Sum & $25.02$ & $2.00$  & $128.4\mathrm{e}{-03}$ & - \\ 
    No-Max-Sum & $24.81$ & $2.32$  & $\mathbf{93.1{{e}}{-03}}$ & -\\ 
    Cond-Max-Sum & $\mathbf{25.18}$ & $\mathbf{1.48}$  & $98.6\mathrm{e}{-03}$ & -\\ 
    MOBIL (Baseline) & $24.80$ & $2.53$  & $176.8\mathrm{e}{-03}$ & -\\ 
    \bottomrule
  \toprule
     Flow:$10000 \,veh/h$& \textit{$v^{avg}_x$ ($m/s$)} &\textit{$v^{avg}_{dev}$ ($m/s$)}  & \textit{$j^{avg}_y$ ($m/s^3 $)} & \textit{$TTS$ ($h$)} \\ 
    \midrule
    Max-Sum & $29.04$ & $1.71$  & $145.0\mathrm{e}{-3}$ & $191.71$\\ 
    No-Max-Sum & $29.00$ & $1.74$  & $\mathbf{98.5{e}{-3}}$ & $191.96$\\ 
    Cond-Max-Sum  & $\mathbf{29.09}$ & $\mathbf{1.64}$  & $127.8\mathrm{e}{-3}$ & $\mathbf{191.39}$\\ 
    MOBIL (Baseline) & $28.68$ & $2.13$  & $162.5\mathrm{e}{-3}$ & $193.94$\\ 
    \bottomrule
    \toprule
    Flow:$15000 \, veh/h$& \textit{$v^{avg}_x$ ($m/s$)} &\textit{$v^{avg}_{dev}$ ($m/s$)}  & \textit{$j^{avg}_y$ ($m/s^3 $)} & \textit{$TTS$ ($h$)} \\     
    \midrule
    Max-Sum & $28.42$ & $2.22$  & $86.4\mathrm{e}{-3}$ & $293.61$\\ 
    No-Max-Sum & $28.36$ & $2.29$  & $\mathbf{62.0{e}{-3}}$ & $294.19$\\ 
    Cond-Max-Sum  & $\mathbf{28.44}$ & $\mathbf{2.21}$  & $71.3\mathrm{e}{-3}$ & $\mathbf{293.41}$\\ 
    MOBIL (Baseline) & $28.04$ & $2.63$  & $87.2\mathrm{e}{-3}$ & $297.37$\\ 
    \bottomrule
  \end{tabular}}
\end{table}

\begin{figure}[t]
    \centering    
    \includegraphics[width=0.45\textwidth]{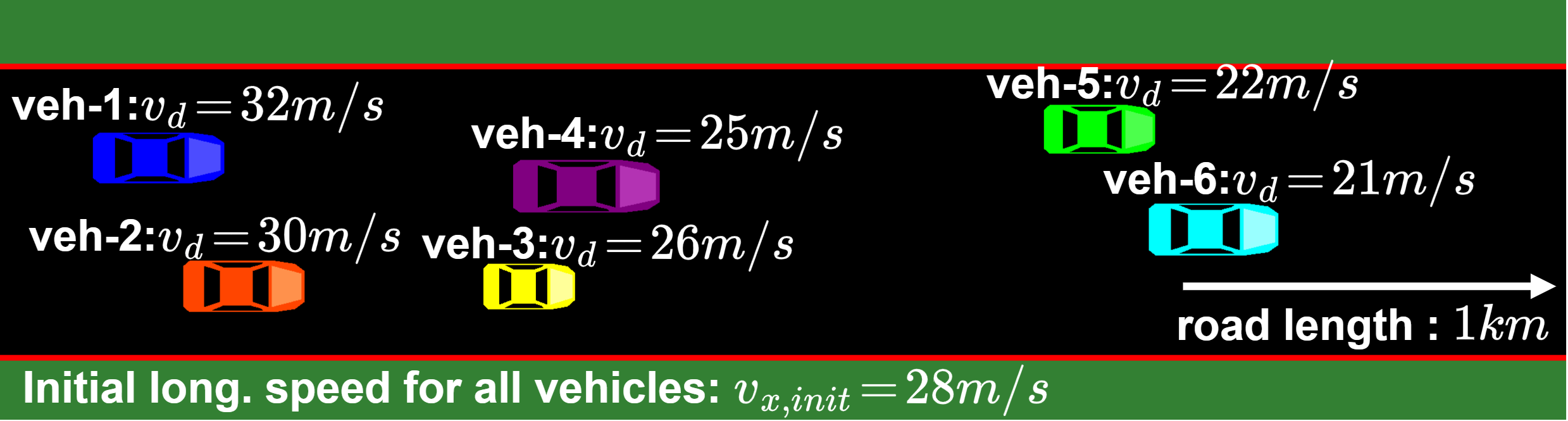}
    \caption{Initial placement of vehicles for the lane-free coordination problem.}    
    \label{fig:small_instance}
    \Description{Snapshot of the initial configuration from the simulation. Alongside the placing, the desired speed goal of each agent is stated.}
\end{figure}

The first type of environment we examine in order to gain empirical insights for our approach is the small experiment with initial placements for vehicles as visualized in Fig.~\ref{fig:small_instance}.
There, we establish three rows where  the desired speeds are set so that vehicles on the back need to overtake the vehicles on the front.
Each vehicle has a different time-estimate for its own next update, meaning that while the vehicles communicate, each one will update its lateral alignment at a different time-step.
The decision regarding the timing of subsequent lateral alignment updates is affected by the proximity of the new selection, resulting again in asynchronous updates due to the independency of agents.
More details on initial conditions and parameter tunings can be found in Appendix~D.2.

\begin{figure}[t]
    \centering
    \includegraphics[width=0.5\textwidth]{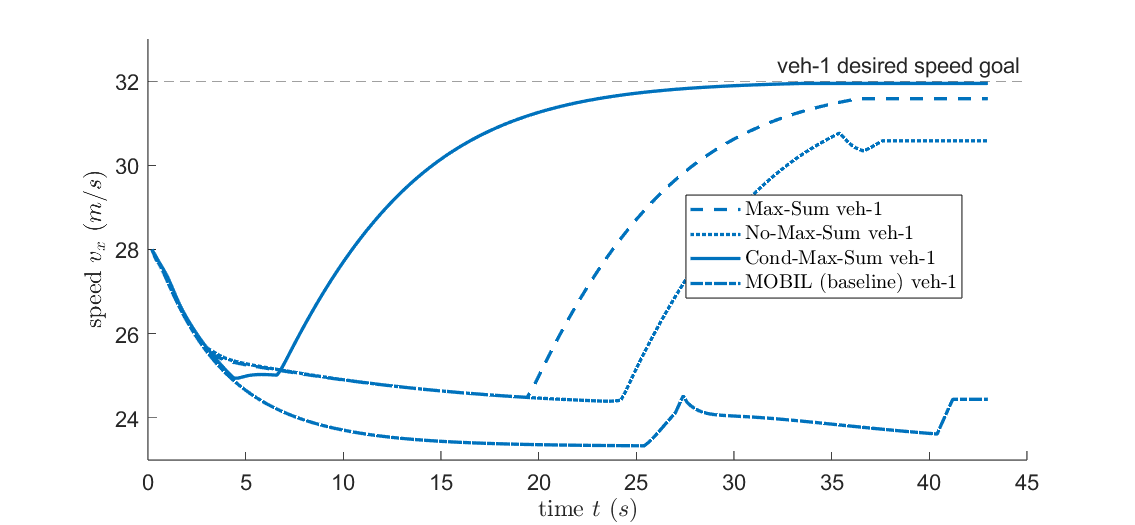}
    \caption{Speed trajectories  of agent veh-1 for all examined methods in the lane-free coordination problem.}
    \label{fig:cond_ms_small_speed}
    \Description{This figure displays the speed trajectories (over time) of agent veh-1. Each examined variant is displayed with different line style. The vehicle manages to reach its desired speed goal significantly faster with Cond-Max-Sum.}
\end{figure}

\begin{figure}[t]
    \centering
    \includegraphics[width=1.0\linewidth]{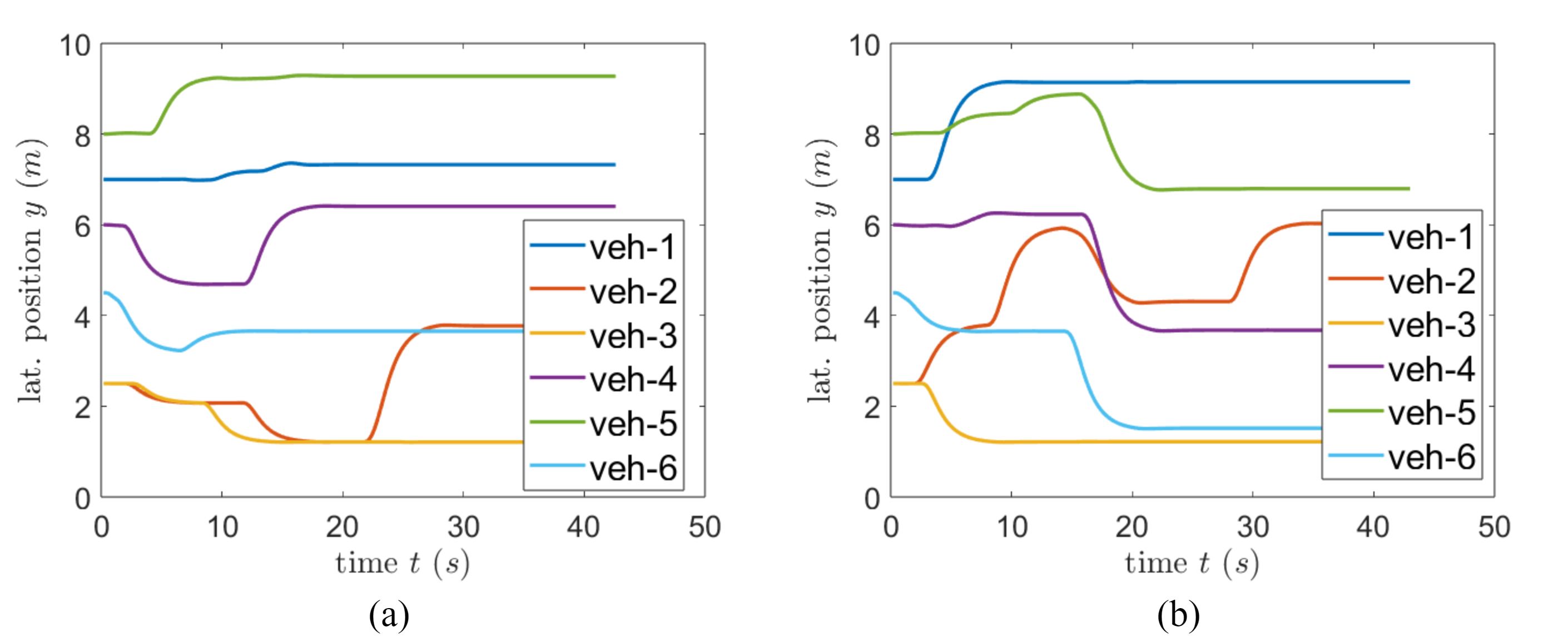}
    \caption{Trajectories of agents' lateral placement for (a) Cond-Max-Sum and (b) Max-Sum, respectively.}
    \label{fig:lat_traj_comp}
    \Description{This figure displays the trajectories (over time) of all 6 agents regarding their lateral placement. Plot lines of each agent are associated based on the color choice. On the left side (a) we visualize the trajectories of Cond-Max-Sum and on the right side (b), the trajectories of the standard Max-Sum. The displayed trajectories show fewer lateral movements on the left sub-figure.}
\end{figure}

Results are presented in Table~\ref{tab:res_large_medium}, where we compare all methods using the above-mentioned metrics.
A more detailed capture of the performance can be found in the longitudinal speed trajectories of Fig.~\ref{fig:cond_ms_small_speed}, where we focus on agent veh-1 located at the beginning of the road, showing how fast and to what extend its desired speed goal is accomplished throughout the simulation time.
Since veh-1 has the highest desired speed, this information---stemming from the regret minimization term of the pairwise factor---is communicated through the message-passing operation.
To put it simply, a vehicle with increased desired speed will result in a heightened regret value when faced with a slower vehicle in front, thereby resulting in a lateral configuration of agents so that veh-1 overtakes.
We can certainly pinpoint Cond-Max-Sum as the superior method when compared either to the standard algorithm or the No-Max-Sum case.
After the initial time period where veh-1 needs to slow down since other vehicles are in front, it then reacts and coordinates more timely with its surroundings, with its desired speed goal being better accommodated without negatively affecting the goals of other agents, as evident in the $v^{avg}_{dev}$ metric in Table~\ref{tab:res_large_medium}.
Then, we can see in Fig~\ref{fig:lat_traj_comp} that Cond-Max-Sum achieves this with fewer lateral maneuvers when compared to the standard algorithm, due to the enhanced message passing.
The value of lateral jerk $j^{avg}_y$ directly provides a measure for the passenger discomfort from these lateral maneuvers.
Even though the No-Max-Sum case has the lowest value of $j^{avg}_y$, when combined with the much inferior speed metrics, it shows that vehicles do not properly harness opportunities for lateral coordination that facilitate overtakes. 
Supplementary results can be found in Appendix~D.3.

\subsection{Open and Large Lane-Free Traffic Settings}\label{subsec:exp_large}

While experiments in a small environment with a specific set of agents allow us to present a detailed depiction of the efficiency by even looking into vehicles' trajectories, we further investigate the use of our distributed framework and Cond-Max-Sum in a much more demanding large-scale open highway environment.
There, new vehicles constantly enter a $2km$ highway's entry point (and are introduced to D-DCOP) while old ones reach the road exit, with traffic flow rates $10000, 15000 veh/h$ that result in hundreds of vehicles populating the road at any given time, specifically around $200,300 veh$ respectively.
There is an induced uncertainty in this type of environment due to the constant emergence of new vehicles (new variables) and frequent changes regarding vehicles' connections (creation/removal of pairwise factors) due to overtakes.
Each vehicle samples a desired speed from a uniform distribution within $\lbrack 25, 35 \rbrack m /s$ upon entrance.
The environment settings for the open highway can be found in Appendix~D.2 and D.4. Then, in~D.5 we include measurements on the size of the FG under such settings and a discussion on the communication overhead per agent.

\balance

The inclusion of a baseline method (MOBIL) without online communication among vehicles serves to further motivate the proposed framework.
As evident in Table~\ref{tab:res_large_medium}, 
all metrics exhibit worse values w.r.t. to any Max-Sum variant, with the exception of jerk (that is close to the standard algorithm for $15000 veh/h$), indicating redundant lateral maneuvers that are not necessarily followed by the intended overtakes.
This shows the effect of the communication among agents when juxtaposed to only observing nearby traffic.

Regarding the three Max-Sum variants in Table~\ref{tab:res_large_medium}, results on the average metrics are consistent with our findings in Sec.~\ref{subsec:exp_small}, albeit with a lower margin on the differences between them. 
This follows from the frequent graph changes in this open environment, which pose an additional challenge compared to the setup in Sec.~\ref{subsec:exp_small} as new pairwise factors can occur unexpectedly and render the calculated messages less compatible under these situations.
Yet, Cond-Max-Sum still consistently outperforms the other variants. 
While the benefit in $v_{dev}^{avg}$ is modest on average, the $TTS$ metric showcases better system level performance.
Then, No-Max-Sum---the variant that is akin to simpler local-search methods (cf. Appendix~A)---always demonstrates inferior speed, but as a consequence, exhibits lower discomfort levels since agents do not explore mutually beneficial decisions to the same extent.
This is also less aligned with the main focus of the factors towards motivating agents to overtake in a coordinated manner whenever desired.

\begin{figure}[t]
    \centering
    \includegraphics[width=0.9\linewidth]{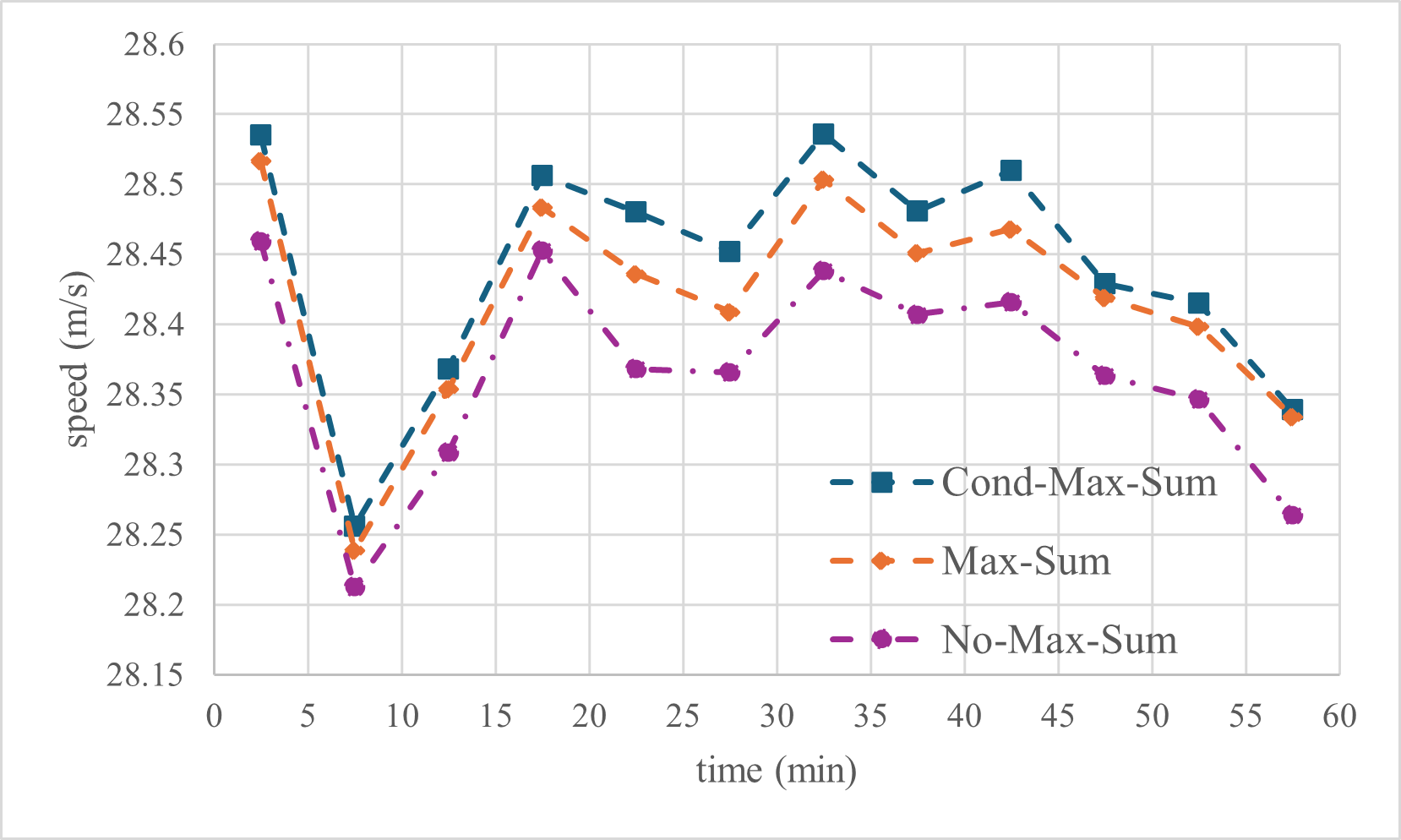}
    \caption{Comparison of measured speed for 5-minute intervals in $15000veh/h$.}
    \label{fig:speed_compare}
    \Description{In this figure, we compare all the speed trajectories of each examined variant for every 5-minute time interval.}
\end{figure}

In order to fully capture the efficiency in the large-scale environment and underline this ``marginal, but consistent'' performance increase for the revised message update of Cond-Max-Sum, we additionally provide Fig.~\ref{fig:speed_compare}, where we divide the $1h$ time span of the simulation to $5 min$ intervals and measure the average speed $v_x^{avg}$ of all agents. 
There, we see how each Max-Sum variant handles the open environment during each time interval of the simulation.
Furthermore, in Appendix D.6, we verify that the results in Fig.~\ref{fig:speed_compare} are statistically significant with $\alpha=0.05$, and contain a (histogram) plot that shows the percentages calculated from all jerk measurements, divided into bins.
For the central bin (ideal case, containing $0 m/s^3$), we have measured $90.8\%, 89.4\%$ and $87.2\%$ for No-Max-Sum, Cond-Max-Sum and Max-Sum, respectively.
With these additional measurements, we can conclude with more confidence that Cond-Max-Sum achieves higher speeds with lower passenger discomfort across all cases, thereby combining two (practically opposing) goals better than the standard Max-Sum update in large-scale settings.

\section{Conclusions and Future Work}\label{sec:concl_fw}

In this work, we proposed a framework for asynchronous decision making in multiagent environments, along with Conditional Max-Sum for enhanced coordination in these settings.
Experimental evaluation in our FG formulation for lane-free traffic exemplifies the applicability of this more realistic framework, along with the enhanced efficiency of Conditional Max-Sum.  
In future work, a natural extension involves environments with {\em external} agents, that is, other entities not complying with the DCOP formulation, and thereby introducing uncertainty to the formulation.
Existing work~\cite{geronymakis2022} already addresses this issue, which can be bundled directly with our approach for a more expansive framework.
Moreover, other algorithms (outside of Max-Sum) in the literature~\cite{dcopsurvey} could be alternatively considered, albeit with possible correspondent adjustments/extensions (as in this case) to accommodate the introduced flexibility.
Finally, in relation to existing work in lane-free traffic, we should point out that to the best of our knowledge this is the only approach that constructs a hybrid system for lane-free environments using safety rules based on dynamic lateral regions for collision avoidance, instead of relying on mathematical models in static environments~\cite{karafyllis_lyapunov} or soft constraints as part of a utility/cost function~\cite{yanumula2023optimalcontrol,troullinos_ijcai}.
Nevertheless, outside of these qualitative differences, direct comparisons could shed light on other trade-offs.


\begin{acks}
The research described in this paper was carried out within the framework of the National Recovery and Resilience Plan Greece 2.0, funded by the European Union - NextGenerationEU (Implementation Body: HFRI. Project name: DEEP-REBAYES. HFRI Project Number 15430).
Moreover, the research leading to these results has received funding from the European Research Council under the European Union’s Horizon 2020 Research and Innovation programme/ERC Grant Agreement n.[833915], project TrafficFluid.
\end{acks}



\bibliographystyle{ACM-Reference-Format} 
\bibliography{cond_ms_aamas25_extended_arxiv}


\newpage
\appendix

\section{Analysis of Message Propagation in Conditional Max-Sum}\label{sec:cond_ms_message_prop}
    
In this section we focus on the effect that the revised update has on the propagation of $q$ messages.
Let us revisit Eq. 4 (from the main text), and decompose the sum over all $q$ messages to $2$ separate sums depending on whether $x_k \in \mathbf{s}_j^{t,e}$ or $x_k \in \mathbf{s}_{j*}^{t-,e}$:

\begin{align}\label{eq:cms_r_t_p_dec}
    r^t_{j \rightarrow i} (x_i) = \max_{\mathbf{s}_j^{t,e}} \Big  \lbrack F^t_j(\mathbf{s}_j) + \sum_{x_k \in \mathbf{s}_j^{t,e}} q^{t-1}_{j \rightarrow k}(x_k) + \sum_{x_k^* \in \mathbf{s}_{j*}^{t-,e}} q^{t-1}_{j \rightarrow k}(x_k^*) \Big \rbrack
\end{align}

\noindent where the summation term with the elements in $\mathbf{s}_{j*}^{t-,e}$ is not affected by the maximization operator. Therefore, we can place it outside the max operator:

\begin{align}\label{eq:cms_r_t_p_dec_out}
    r^t_{j \rightarrow i} (x_i) = \max_{\mathbf{s}_j^{t,e}} \Big  \lbrack F^t_j(\mathbf{s}_j) + \sum_{x_k \in \mathbf{s}_j^{t,e}} q^{t-1}_{j \rightarrow k}(x_k) \Big \rbrack + \sum_{x_k^* \in \mathbf{s}_{j*}^{t-,e}} q^{t-1}_{j \rightarrow k}(x_k^*) 
\end{align}

\noindent Before advancing, we define: 

\begin{eqnarray}\label{eq:r_t_max}
    r^{t,max}_{j \rightarrow i} (x_i) = \max_{\mathbf{s}_j^{t,e}} \Big  \lbrack F^t_j(\mathbf{s}_j) + \sum_{x_k \in \mathbf{s}_j^{t,e}} q^{t-1}_{j \rightarrow k}(x_k) \Big \rbrack 
\end{eqnarray}

\noindent and:
\begin{eqnarray}\label{eq:r_t_star}
    r^{t,*}_{j \rightarrow i} =  \sum_{x_k^* \in \mathbf{s}_{j*}^{t-,e}} q^{t-1}_{j \rightarrow k}(x_k^*) 
\end{eqnarray}

\noindent for reasons of compactness. Note that the sum outside of the max operator does not depend on $x_i$, meaning that it is an added constant value independent of $x_i$. 
Since this value is the same across all different $x_i$ inputs, then we can already deduce that these propagated values do not affect the decision making of agents.
Below, we examine this claim in more mathematical terms.

Applying Equations~\ref{eq:r_t_max}\&~\ref{eq:r_t_star} to~\ref{eq:cms_r_t_p_dec_out}, we have:

\begin{eqnarray}\label{eq:cms_r_t_p_dec_comp}
    r^t_{j \rightarrow i} (x_i) = r^{t,max}_{j \rightarrow i} (x_i) + r^{t,*}_{j \rightarrow i} 
\end{eqnarray}

\noindent Then, applying Eq.~\ref{eq:cms_r_t_p_dec_comp} to the $q$ messages update (Eq.~2 from the main text):

\begin{eqnarray}\label{eq:q_ij_t_a}
    q^t_{i \rightarrow j}(x_i) &=& c_{ij} + \sum_{k \in M_i \setminus j} \Big  \lbrack r^{t,max}_{j \rightarrow i} (x_i) + r^{t,*}_{j \rightarrow i}  \Big  \rbrack \nonumber \\
    q^t_{i \rightarrow j}(x_i) &=& c_{ij} + \sum_{k \in M_i \setminus j} \Big  \lbrack r^{t,max}_{j \rightarrow i} (x_i) \Big  \rbrack + (|M_i|-1)\cdot r^{t,*}_{j \rightarrow i}
\end{eqnarray}

Since the normalization constant is calculated to satisfy:\\ $\sum_{x_i} q^t_{i \rightarrow j} (x_i) = 0$ in cyclic graphs, then solving for $c_{ij}$, we get:

\begin{eqnarray}\label{eq:c_ij_cond}
    c_{ij} &=& -\frac{1}{|X_i|} \cdot \sum_{x_i} \Big  \lbrack \sum_{k \in M_i \setminus j} \big  \lbrack r^{t,max}_{j \rightarrow i} (x_i)  \big  \rbrack  + (|M_i|-1)\cdot r^{t,*}_{j \rightarrow i}\Big  \rbrack \nonumber \\
    c_{ij} &=& -\frac{1}{|X_i|} \cdot \bigg \lbrack \sum_{x_i}  \Big  \lbrack \sum_{k \in M_i \setminus j} \big  \lbrack r^{t,max}_{j \rightarrow i} (x_i)\big  \rbrack \Big  \rbrack  + \nonumber \\ &+& |X_i|\cdot (|M_i|-1)\cdot r^{t,*}_{j \rightarrow i} \bigg \rbrack  \nonumber \\
    c_{ij} &=& -\frac{1}{|X_i|} \cdot \bigg \lbrack \sum_{x_i}  \sum_{k \in M_i \setminus j} r^{t,max}_{j \rightarrow i} (x_i) \bigg \rbrack  - (|M_i|-1)\cdot r^{t,*}_{j \rightarrow i} 
\end{eqnarray}

\noindent where $|X_i|$ is the number of discrete values for $x_i$, meaning that the constant added term $r^{t,*}_{j \rightarrow i}$ on $r^t_{j\rightarrow i}$ messages is in fact never communicated to other agents.
Moreover, it does not affect the decision update of the agent either since it is constant: 

\begin{eqnarray}
    x_i^* &=& \arg \max_{x_i} \sum_{j \in M_i} r^t_{j \rightarrow i} (x_i) \nonumber \\
    &=& \arg \max_{x_i} \sum_{j \in M_i} \lbrack r^{t,max}_{j \rightarrow i} (x_i) + r^{t,*}_{j \rightarrow i} \rbrack \nonumber \\
    &=& \arg \max_{x_i} \big \lbrack \sum_{j \in M_i} r^{t,max}_{j \rightarrow i} (x_i) + \sum_{j \in M_i} r^{t,*}_{j \rightarrow i} \big \rbrack \nonumber \\
    &=& \arg \max_{x_i} \sum_{j \in M_i} r^{t,max}_{j \rightarrow i} (x_i)
\end{eqnarray}

In practice, we can disregard the $q$ messages sent from agents not complying with the time estimate criterion, without affecting the result.
As such, in the case of factors connecting two agents, the value $q^{t-1}_{k\rightarrow j}(x_k^*)$ could be omitted from Eq.3 (from the main text).
Therefore, when the agent chooses not to maximize, it essentially bounds the associated factor's input element with the sent value for $x_k^*$, and locally maximizes over the connected factors when it updates its decision on $x_i^*$.
This reflects distributed local search algorithms in the context of DCOPs, specifically MGM~\cite{mgm_paper}, which is the non-stochastic variant of the popular baseline DSA~\cite{dsa_paper} algorithm for DCOPs.
We examine the efficacy of this in our experimental evaluation through the No-Max-Sum variant, where the agents never maximize over the others' variables and only rely on the broadcasted assignments.

\section{Lane-Free Traffic Environments with Dynamic Lateral Regions}\label{lf:d_lr}

In this section we present in detail our formulation of lane-free environments that enables lane-free lateral operations through a Dynamic DCOP.
The primary tool that we design to enable coordination in a manner suitable for strategic coordination in lane-free traffic is that of {\em dynamic lateral regions}.
With this, the vehicles can interpret the observed and/or communicated information from nearby traffic in a way that provides a {\em real-time structured representation} of the environment, and decide upon low-level operations regarding their acceleration and account for safety.

\subsection{Formation of Lateral Regions}\label{subsec:lat_regions_form}
\begin{figure}[t]
    \centering
    \includegraphics[width=0.48\textwidth]{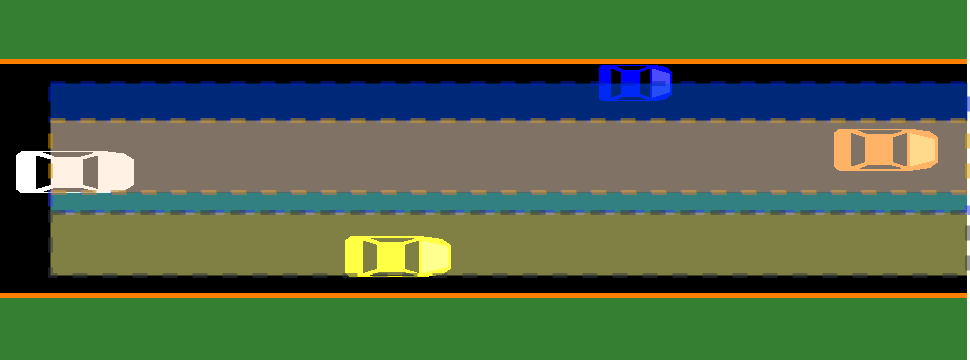}
    \caption{Formed lateral regions from downstream traffic the perspective of agent $i$.}
    \label{fig:lat_regions_downstream}
    \Description{Visual illustration of the lateral regions' formation due to traffic located in front of the agent/vehicle we focus at.}
\end{figure}

\begin{figure}[b]
    \centering
    \includegraphics[width=0.48\textwidth]{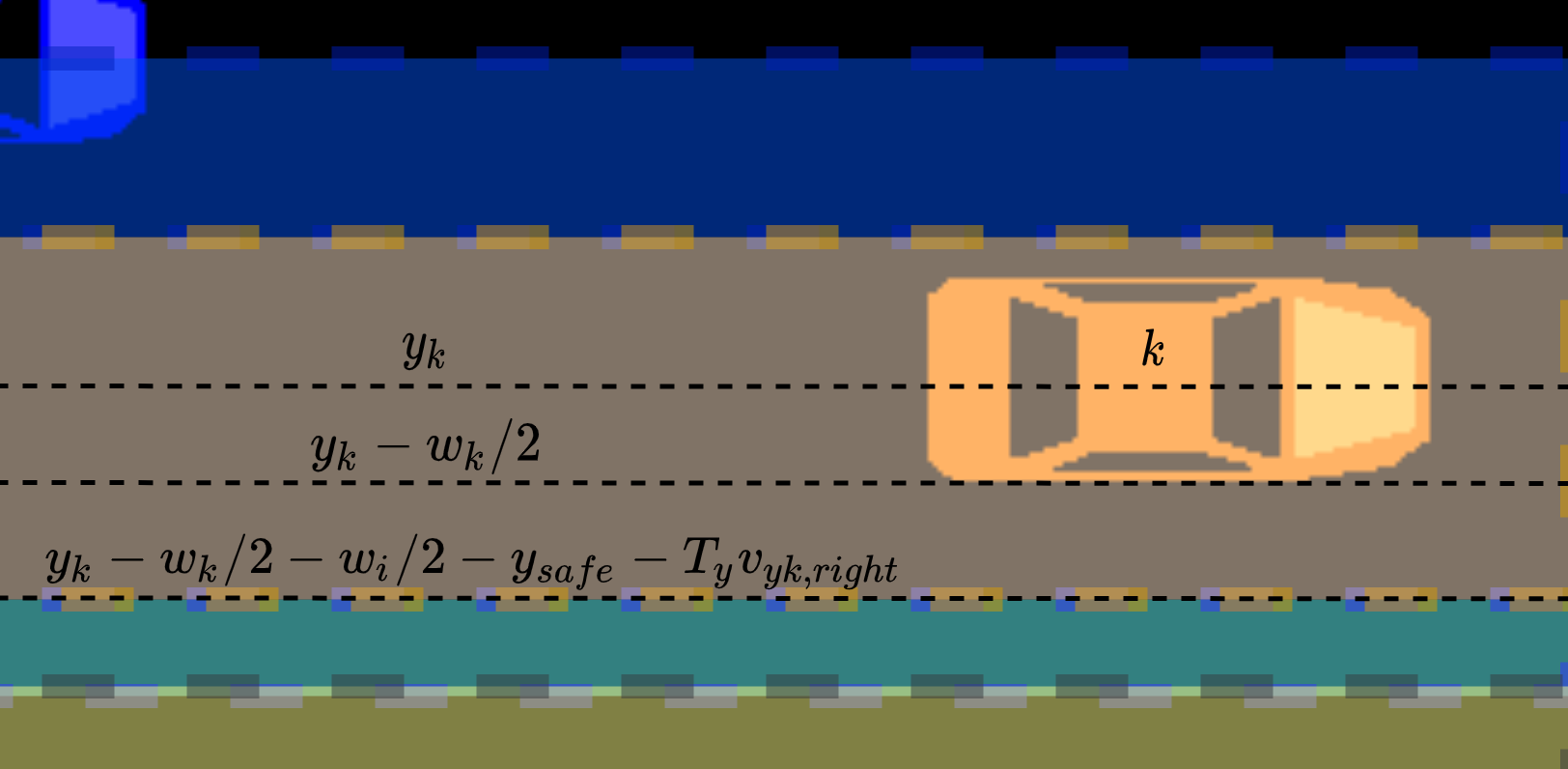}
    \caption{Example of lateral region formation based on the lateral information of observed vehicles.}
    \label{fig:lat_regions_detail}
    \Description{Detailed view of the parameters that affect the reach of each region.}
\end{figure}

This process initiates from the perspective of a single ego vehicle $i$. First, we distinguish between upstream (vehicles on the back of $i$) and downstream traffic (vehicles on the front of $i$).
For downstream traffic, we scan all vehicles according to a specified observation distance ${obs}_x$ longitudinally (in front), and construct the lateral region that each foreign vehicle introduces as illustrated in Fig.~\ref{fig:lat_regions_downstream}.
The partitioning of the regions is done based on the monitored information from each observed vehicle $k$, namely its: lateral placement on the road $y_k$, lateral speed $v_{y,k}$ and physical dimensions on the lateral axis, i.e., its width $w_k$.
This is visually shown in Fig.~\ref{fig:lat_regions_detail}, where from the perspective of the white ego vehicle ($i$), a lateral region due to $k$ is formed according to this information, and accounting for safety related gaps, a distance $y_{safe}$, and a time-gap $T_y$ that is only added when the vehicle has a lateral speed towards this direction: $v_{vk,right}= -v_{y,k}\cdot (v_{y,k}<0)$, and : $v_{vk,left}= v_{y,k}\cdot (v_{y,k}>0)$.\footnote{Positive lateral speed indicates movement towards the left direction, while negative values that the vehicle is moving right.} 
The constructed regions correspond to where the center point of the ego vehicle can be positioned, hence the reason we see the utmost left and right regions not reaching the road's physical boundaries as they take into account the ego's width.

At any time, the ego vehicle is located at a specific lateral region, in which we consider that its longitudinal behaviour is being influenced by the front vehicle occupying this region.
As such, the longitudinal control (gas/brake) of the vehicle can be decided according to a car-following method as done typically in lane-based environments, with the vehicle in front as the leader to follow.
For this task, we employ the Enhanced Intelligent Driver Model (EIDM)~\cite{kesting2010enhanced}, which is an extension of one of the most popular car-following methods, that calculates the acceleration of the vehicle $i$ on the longitudinal axis (corresponding to a gas/brake response), taking into account its desired speed $v_i^d$ while respecting a desired time-gap value $T_x$ with the vehicle in front to avoid critical situations.
In this manner, given two vehicles $(i,j)$ with $j$ being in front of $i$, we can calculate the acceleration $a_{i, j}$ of $i$ when $j$ is in front according to EIDM.

Likewise, we can calculate this acceleration for each lateral region, meaning we can have an estimate of the acceleration of ego vehicle depending on its lateral placement.
These acceleration estimates are instrumental in our approach, as they quantify the {\em value} of residing at a lateral region, and consequently the {\em benefit} of shifting laterally to a different region by simply comparing the corresponding acceleration evaluations.
The calculated accelerations also provide information about the criticality of any selected positioning, e.g., a negative value indicating strong braking will be calculated for a lateral region containing a vehicle being too close or/and operating at a significantly lower speed than ego.
At the same time, the acceleration values solely reflect our vehicle's desired speed goal whenever there is no need for gap maintenance with the front vehicle.

An important characteristic in lane-free traffic is that of nudging, meaning that vehicles' behaviour can be influenced by surrounding vehicles located upstream.
We can directly incorporate nudging for this formulation by constructing a second set of lateral regions now from upstream traffic by following exactly the same logic. 
Then, a ``follower'' vehicle $k$ can be additionally specified according to the ego's position w.r.t. to the formed lateral regions from the vehicles in the back, as shown in Fig.~\ref{fig:lat_regions_viz}.
Finally, we can similarly obtain estimates for the acceleration but now from the perspective of the vehicle on the back wrt to our ego, i.e., $a_{k,i}$.
From this perspective, the acceleration associated with each lateral region provides us with how the ego's  lateral alignment affects these vehicles, e.g., a significant negative response indicates that the vehicle on the back is either too close or attempts to overtake.

Note that the regions introduced by each vehicle can occupy the same lateral space. 
For this, we give priority to the vehicle with the lowest acceleration estimate (which is typically the one closest to $i$), e.g., in Fig.~\ref{fig:lat_regions_viz}, the blue region by vehicle $l$ occupies lateral space that would be otherwise part of $m$'s lateral region.

\begin{figure}[b]
    \centering
    \includegraphics[width=0.48\textwidth]{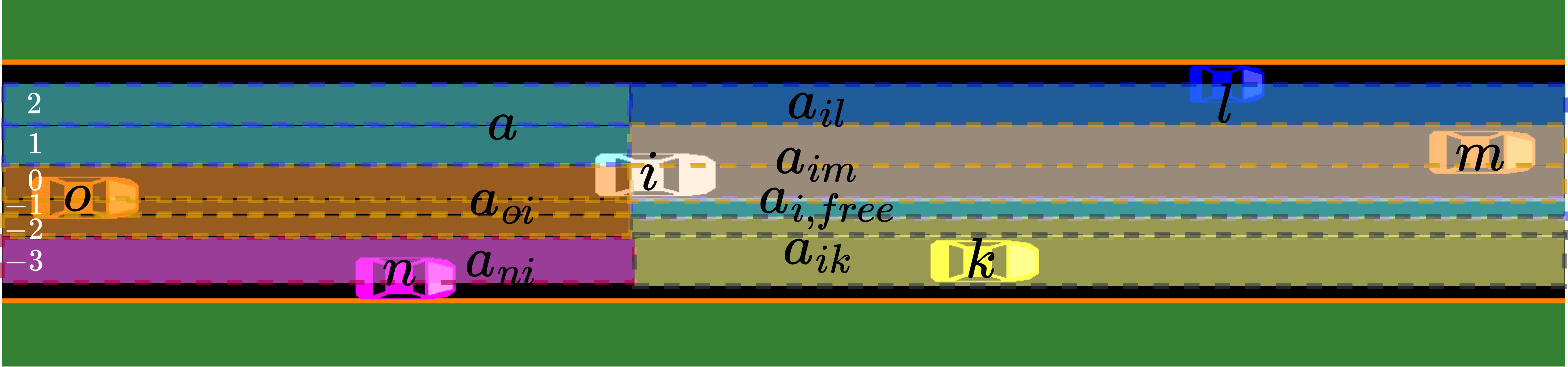}
    \caption{Formed lateral regions partitioned by downstream and upstream traffic for agent $i$.}
    \label{fig:lat_regions_part}
    \Description{Visual illustration of the lateral regions' formation due to surrounding traffic (front and back) from the perspective the agent/vehicle we focus at.}
\end{figure}

\subsection{Regulation of Lateral Alignment in Lane-Free Traffic}\label{subsec:reg_lat_align}

With the formulation of lateral regions as described above, we ``filter out'' potentially unsafe decisions by examining the feasibility of a lateral shift from one lateral alignment to the next. 
In order to rely on a common indexing $r$ for downstream and upstream traffic, we further partition downstream and upstream lateral regions, as shown in Fig.~\ref{fig:lat_regions_part} with the white colored numbers at the left indicating the indexing of the respective region.
Given the current lateral position $y_i$ of ego vehicle and a desired lateral alignment $y_i^*$ from a policy, we can directly determine the desired lateral region 
$r^*$ for that position and assign the closest {\em safe} region $r_d$ that conforms to the following safety criterion.

For every lateral region $r$ between the current region $0$ and the desired one $r^*$, we examine whether the shift is feasible based on the acceleration estimate for the vehicles upstream $a_{\cdot,i}$ and downstream $a_{i,\cdot}$.
This establishes that the lateral shift will not cause a critical situation with the vehicles on that direction.
The lateral shift towards a desired region is considered safe only if the following condition for a {\em safe} deceleration value $b_{safe}$ is satisfied for all intermediate regions, i.e., $a_{\cdot,i} \geq -b_{safe}$.
For instance, if $i$ in Fig.~\ref{fig:lat_regions_part} wishes to laterally shift towards region $-3$, then we require that both $a_{o,i}\geq -b_{safe}$ and $a_{n,i}\geq -b_{safe}$, along with $a_{i,k}\geq -b_{safe}$.
If this condition is not met, then we select the region closest to the initial goal that satisfies it.
As such, we obtain a {\em safe} lateral region $r_d$ after this procedure (worst case scenario is that $r_d=0$, i.e., we can only move laterally within the limits of our current region).
Then, the actual desired lateral positioning is calculated as:

\begin{equation}
    y_i^d = \max {\big \{ \min {\{ y_i^{*},y_h(r_d)-y_{thr}\}}, y_l(r_d) + y_{thr} \big \}}
\end{equation}

\noindent where $y_h(r_d),y_l(r_d)$ are the highest and lowest lateral point of the selected lateral region $r_d$, and $y_{thr}$ is a small threshold value so that the vehicles do not reside at the upper or lower limit of the region.
This safety rule can be viewed as a lane-free extension of the one employed in lane-based environments, where an intended lane-change can be aborted if the estimated acceleration at the destination lane does not comply with a desired deceleration value ($b_{safe}$).

\subsection{Control input of vehicles}

We now have the information on the lateral regions of an ego vehicle $i$, and its desired lateral placement $y_i^d$.
In order to control the vehicle, we need to translate this to two acceleration values $a^x_i,a^y_i$ in $m / s^2$.

For the longitudinal acceleration (gas/brake pedal), we simply determine the ``leader'' vehicle in front of us, along with the ``follower'' depending on the current lateral region the ego occupies (downstream and upstream), e.g., for region $0$ in Fig.~\ref{fig:lat_regions_part}, leader is vehicle $m$ while follower is $o$.
The final acceleration value is calculated through the EIDM model, but with a significant difference that incorporates nudging behaviour, i.e., the influence of the vehicle on the back as well.
Such extensions on IDM are also investigated in related work~\citep{YI2022127606} for lane-based traffic.
Especially in the lane-free traffic domain, the influence of nudging is quite significant~\citep{lane_free_journal}, and has a substantial impact on preliminary empirical results with our approach as well.
Therefore, the longitudinal acceleration $a^x_i$ of each vehicle $i$ with leader vehicle $j$ and vehicle $k$ on the back is calculated as:

\begin{equation}
    a^x_i = a_{i,j}+ \alpha \cdot \gamma \frac{s^*_{ki}}{s_{ki}}^2 
\end{equation}

\noindent where the second term contains the ``reaction'' of $i$ due to the presence of vehicle $k$ behind.
This term is directly taken by the IDM model, where $s^*_{ki}$ is the desired longitudinal distance between $k$ and $i$ according to a provided time-gap value $T_x$, and $s_{ki}$ is the monitored one.

For instance, if $k$ intends to decelerate due to $i$'s presence, this term will also cause $i$ to accelerate, therefore mitigating the reaction of $k$.

For the lateral acceleration $a_i^y$ (left/right movement), we calculate based on a simple PD controller~\cite{kuo1995automatic} that drives the vehicle towards its desired lateral alignment $y_i^d$ while avoiding unnecessary oscillatory behaviour. This is accomplished with the following form:

\begin{equation}\label{eq:ay_update}
    a^y_i = K_p \cdot (y_i^d - y_i) + K_d \cdot (-v_{y,i})
\end{equation}

\noindent where $K_p,K_d$ are the proportional and derivative gains respectively, calibrating how fast and smoothly the vehicle reacts.
Essentially, the PD controller automates how the vehicle actually targets the desired lateral position $y_i^d$ in a continuous space without causing unstable behaviour.

\section{Multiagent Coordination in Lane-Free Traffic (Supplementary Material)}\label{sec:mas_lf2}

\subsection{Pairwise Factors (with Supplementary Material)}\label{subsec:fg_lf2}

The FG formulation contains a second type of a pairwise factor $F_{p}(x_i,x_j)$ that connects two vehicles $i$ and $j$, with $j$ preceding $i$. Its presence serves to motivate both $i$ and $j$ at moving laterally according to $i$'s {\em desire} to overtake through {\em regret minimization}.
Notably, since the factor affects both $i$ and $j$'s decision due to their involvement, they can accordingly control their lateral behaviour in a coordinated manner.
This is accomplished by the following formulation:

\begin{align}\label{eq:f_ij_s}
    F_{p}(x_i,x_j) = regret(x_i,x_j) + comfort(x_i,x_j)
\end{align}

\noindent where the first term is the calculated {\em regret} from the perspective of the receding agent $i$, and has the following form:

\begin{equation}\label{eq:regret_s}
    regret(x_i,x_j) = -R_c\cdot (a_{i,free} - a_{i, j})^2 \cdot overlap_{ij}(x_i,x_j)
\end{equation}

\noindent and the value $a_{i,free}$ is the calculated acceleration of $i$ when located in lateral regions without a leader, meaning that this value only accounts for the desired speed objective of the agent and that $a_{i,free}\geq a_{i,j}$.
As such, the difference $(a_{i,free} - a_{i, j})$ expresses the {\em regret} of agent $i$ for having $j$ {\em in front} of it,\footnote{$i$ observing $j$ as leader for the examined configuration $y_i',y_j'$.} using a positive coefficient $R_c$ and a negative sign to comply with the algorithm's maximization criterion.
Note that whenever this type of factor connects two vehicles, this regret value is assigned to it only for configurations of $x_i,x_j$ that would result in the agents having their lateral alignments ``overlap'' at any point during lateral deviation from the current placement $y_i,y_j$ towards the examined one $y_i',y_j'$, i.e.:

\begin{align}
    overlap_{ij}(x_i,x_j) = 
    \begin{cases}
        1 & \text{if }  \; inRange_{ij}(y_i, y_j, y_i',y_j')\\
        0.75 & \text{if } onlyOverlap_{ij}(y_i, y_j, y_i',y_j')\\
        0    & \text{otherwise.}
    \end{cases}
\end{align}

\noindent where the $inRange_{ij}$ operator checks whether the desired configuration $y_i',y_j'$ results in the two vehicles observing one another in the same lateral region when the vehicles have reached  $y_i', y_j'$.
Instead, a smaller percentage of the regret is assigned if the examined combination of lateral alignments $y_i', y_j'$ does not result in the vehicles at the same lateral region, but they still have to laterally overlap one another, e.g., if initially $j$ is right of $i$ but the examined configuration results in $j$ being left of $i$.
If neither of the above holds true, then no regret is assigned to the factor.

\subsection{Formation of Pairwise Connections}\label{subsec:form_pairwise}

Each agent is associated with one single factor, as illustrated in Fig.~\ref{fig:connected_vehs}.
A pairwise factor $F_{p}(x_i,x_j)$ between $i,j$ is formed when the agents can observe one another based on the observational distance ${obs}_x$, and according to their lateral distance $dy$, based on the following condition:

\begin{equation}
    dy \leq C_{range}\cdot y_{range} + y_{safe}
\end{equation}

\noindent where $dy$ is the lateral distance of the vehicles accounting for their respective widths, and $y_{range}$ is the selected range of lateral deviation for the control variables $x_i$.
For instance, if $y_{range}=3m$, then the lateral deviation takes values within the range $x_i=[-3m,3m]$, resulting in each vehicle being able to move towards the left or right direction by $3$ meters with respect to its current lateral position.
Then, $C_{range}$ is a positive coefficient, and $y_{safe}$ is the safety distance gap also used for the design of lateral regions (see Sec.~\ref{subsec:lat_regions_form} in this document).

Additionally, each agent $i$ prunes its own connections with other agents in order to conform to upper limits $N_{front}^{max}, N_{back}^{max}$ regarding the number of pairwise connections it can have downstream and upstream, respectively. 
The pruning operation for agent $i$ is performed by ordering the candidate pairwise factors $F_{p}(x_i,x_j)$ downstream according to the acceleration values $a_{i, j}$ in ascending order, and only forming the first $N_{front}^{max}$ connections.
The same procedure is followed for upstream connections $F_{p}(x_j,x_i)$, but now according to the acceleration values $a_{j,i}$.
Note that each agent performs this operation from its own perspective of the problem, and each pairwise factor is formed only if both agents agree to have this connection after the pruning process. 
This procedure is part of line 2 in Alg.1 of the main text.

\subsection{Time-estimates update (Supplementary}\label{sec:async_max_sum_lf2}

The underlying movement dynamics of the vehicles in our case is the double-integrator model, meaning that at every new time-step $t$, the lateral position $y_i$ of the vehicle is updated as:

\begin{equation}\label{eq:y_i}
    y_i = y_i + v_{y,i}\cdot dt + 0.5\cdot a^y_i \cdot dt^2
\end{equation}

\noindent and:

\begin{equation}\label{eq:v_yi}
    v_{y,i} = v_{y,i} + a^y_i\cdot dt
\end{equation}

\noindent where $dt$ is the discrete time-step length for the simulation environment.
The PD controller in Eq.~\ref{eq:ay_update} contains the lateral position $y_i$ and speed $v_{y,i}$ of the vehicle. 
Therefore, we can directly predict in how many time-steps the desired lateral positioning $y_i^d$ will be reached since we know the model of the vehicle's movement dynamics.
More specifically, we iteratively apply Eqs.~\ref{eq:y_i},~\ref{eq:v_yi} and Eq.~\ref{eq:ay_update} until the desired position is reached within a small $y_e$ distance, i.e., $|y_i^d - y_i | \leq y_{e}$.

\section{Experimental Evaluation}\label{sec:exp_eval2}

\subsection{MOBIL as a baseline method in Lane-Free Traffic}

MOBIL~\cite{kesting2007mobil} is one of the most well-known  methods that provide a policy for lane-changing behaviour.
MOBIL models how humans would decide to change lanes, with parameters that calibrate its behaviour regarding the vehicle's eagerness to change lane whenever an opportunity on an adjacent lane arises; or how polite it is to upstream traffic, i.e., the vehicle performing a lane change not necessarily for its own benefit, but taking into account how this lane-change would affect the vehicles located upstream on the current and the candidate lane.
This notion can be directly translated to lane-free traffic environments by replacing the adjacent lanes with adjacent lateral regions of the vehicle. 
Given a candidate lateral region $r$ and the currently residing lateral region $r_0$, we examine the potential lateral shift from $r_0$ to $r$ with the following condition:

\begin{equation}\label{eq:mobil}
    a_i(r) - a_i(r_0) > p [(a_u^{-i}(r)-a_u^i(r))+(a_u^i(r_0)-a_u^{-i}(r_0))]+ a_{thr}
\end{equation}

\noindent where $a_i(r), a_i(r_0)$ are the acceleration values for ego vehicle $i$ on the candidate $r$ and current $r_0$ lateral regions respectively. Then, $a_u^i(r),a_u^{-i}(r)$ are the acceleration estimates for the vehicle $u_r$ located at the candidate region $r$ upstream of $i$, accounting for the two possible outcomes: $a_u^i(r)$ corresponds to $i$ performing the lateral shift, meaning that $u_r$'s acceleration will be influenced by $i$; and $a_u^{-i}(r)$ corresponds to $i$ remaining at the current region $r_0$, meaning that $u_r$ will have the current vehicle in front.
As such, the term $(a_u^{-i}(r)-a_u^i(r))$ is the acceleration impact on vehicle $u_r$ located at the candidate region $r$ if the lateral shift takes place.
Likewise, $a_u^i(r_0),a_u^{-i}(r_0)$ expresses the same estimates, but for the vehicle $u_{r_0}$ located upstream of $i$ at the current lateral region $r_0$. 
Therefore, the term $(a_u^i(r_0)-a_u^{-i}(r_0))$ is the impact of the lateral shift of $i$, but for the vehicle $u_{r_0}$.

Moreover, the influence of these two terms regarding $i$'s decision is calibrated with a {\em politeness} coefficient $p$.
Typical values of $p$ are within the range $[0,1]$, with $0$ expressing that vehicle $i$ is not influenced by the impact of its lateral shift for the affected upstream vehicles, and $1$ the exact opposite, namely that vehicle $i$ will perform the lateral shift only if the total benefit (its own and for the two vehicles upstream $u_r,u_{r_0}$) is positive.
Finally, there is a non-negative threshold value $a_{thr}$ calibrating the minimum benefit $i$ should observe in order to perform the lateral shift.
This is important because small values of $a_{thr}$ can cause the vehicle to have quite oscillatory behaviour, and results in nearby traffic to respond to such situations.

In a conventional lane-based environment, we would examine the two adjacent lanes of the vehicle (or one if the vehicle is next to a boundary of the road), and perform the lane-change if the condition in Eq.~\ref{eq:mobil} is met for one of them, and the safety rule (see Sec.~\ref{subsec:reg_lat_align}) regarding $b_{safe}$ is satisfied.
We extend the same principle in lane-free settings, but instead of checking only adjacent regions, we use the $y_{range}$ distance\footnote{This is the same distance that we use for the factor graph formulation in order to properly compare the different methods.} to indicate how far the vehicle can search for regions (left or right) that have an acceleration benefit, and select the region with the highest benefit that satisfies the condition along with the safety rule.

\begin{table}[t]
  \centering  
  \caption{Parameter Tuning for Factor Graph \& Max-Sum Formulation}  
  \begin{tabular}{|c|c||c|c|}  
  \hline
     $N^{max}_{front}$ & 6 & $T_{min}$ & $4 sec$\\ 
    \hline
    $N^{max}_{back}$ & 6 & $T_{max}$ & $6 sec$\\ \hline
    $y_{range}$ & 3.5 m & $t_e$  & $1 sec$ \\ \hline
    $|X_i|$ & 15 & $y_e$  & $0.01 m$\\ \hline
    $R_c$ & 5 & $C_{range}$& $1.25$\\ \hline
    $C_c$ & 0.05 &  $B_c$ & 12 \\ \hline    
  \end{tabular}
  \label{tab:fg}  
\end{table}

\subsection{Technical Details}\label{subsec:tech}

All experiments are conducted with the TrafficFluid-Sim~\cite{tsim_sumo2022} lane-free microscopic simulator, an extension of SUMO~\cite{sumopaper} appropriate for lane-free environments.
The codebase was developed in C++, and since this framework operates on a single iteration per time-step, execution times are quite small as we later report for the largest setting.
Upon acceptance, we will share a proper version of the codebase with usage instructions.
Across all simulations, a discrete time-step of $dt=0.2sec$ is used.

Here, we contain information regarding the parameter tunings that are common across all scenarios.
First, in Tab.~\ref{tab:fg}, we provide the parameter settings relevant to the Factor Graph formulation and the Conditional Max-Sum algorithm, where $|X_i|$ is the number of discrete values within the range $[-y_{range},y_{range}]$ for variables $x_i$.
Then, in Tab.~\ref{tab:lr} we include the parameter settings relevant to the formation of lateral regions, EIDM for the longitudinal behaviour of vehicles and the acceleration estimates for the formulation of lateral regions, along with the tuning for MOBIL.
There, $a_{max}$ is the maximum acceleration (longitudinal, gas behaviour)  that the vehicles can have, and $b_{safe}$ is the value used both for EIDM, and for the safety rule that regulates the vehicles' lateral movement.
Moreover, $a_{severe}$ is the minimum allowed acceleration value (longitudinal, brake behaviour).
Vehicles with such low acceleration values indicate a ``severe'' situation (i.e., the need to react urgently in order to avoid a collision).

Additionally, $T_x$ is the time-gap value relevant to the longitudinal behaviour of vehicles, and $T_y$ is used for the lateral regions formation (cf Fig.~\ref{fig:lat_regions_detail}).
The simulation environment is discrete-time, but we report on the time-related information in seconds instead of number of time-steps.
For instance, the value of $T_{min}=4sec$ corresponds to $4/0.2=20$ time-steps.
Finally, ${obs}_x$ is the observational distance of each vehicle, based on which it monitors nearby traffic downstream and upstream.

The baseline MOBIL method is tuned with $a_{thr}=0.8$ and politeness $p=0.5$ after preliminary empirical investigations. We selected the values that better balanced jerk behaviour and the desired speed objective.
Without careful tuning, MOBIL can exhibit quite intense jerk behaviour especially for more aggressive settings ($a_{thr},p$ values close to $0$) due to the dynamic nature of the problem and the lack of communication.
At the same time, we never observed significant improvement in terms of desired speed objective, especially in intense settings.

\begin{table}[b]
  \centering
  
  \caption{Parameter Tuning relevant to Lateral Regions, EIDM, and MOBIL} 
  \begin{tabular}{|c|c||c|c|}  
  \hline
     $a_{max}$ & $1.5 m/s^3$ & $a_{thr}$ & $0.8$\\ 
    \hline
    $b_{safe}$ & $2 m/s^3$ & $p$ & $0.5$\\ \hline
    $a_{severe}$ & $-5 m/s^3$ &   $x_{safe}$ & $0.3m$\\ \hline
    $T_x$ & $0.4 sec$ & $y_{safe}$ & $0.2m$ \\ \hline
    $T_y$ & $0.4 sec$ &  $\gamma$ & $0.7$  \\ \hline
    ${obs}_x$ & $30m$ & & \\ \hline   
  \end{tabular}
  \label{tab:lr}  
\end{table}

\begin{figure*}[t]
    \centering
    \includegraphics[width=1\textwidth]{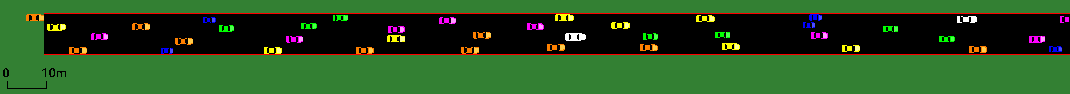}
    \caption{Snapshot of the first $250m$ of the open highway.}
    \label{fig:highway_open}
    \Description{Top-down view of the first 250 meters of the open highway, populated with lane-free vehicles.}
\end{figure*}

\subsection{Lane-Free Coordination Problem}

In this section, we include additional details for the initialization of the problem (as visualized in Fig.~5 of the main text).
All simulations initiate from time $0 sec$, with each vehicle having a different initialization regarding its previous update, consequently affecting its time window $[T_{min}=4sec, T_{max}=6sec]$.
For instance, veh-1 has an initialization of $-1 sec$, meaning that its first time window to consider updating its variable is $[3sec,5sec]$.
Each vehicle's initial timing is reported in Tab.~\ref{tab:init_time}.
The examined highway has a length of $1km$ and a width of $10.2m$, corresponding to a conventional 3-lane highway.

\begin{figure}
    \centering
    \includegraphics[width=0.45\textwidth]{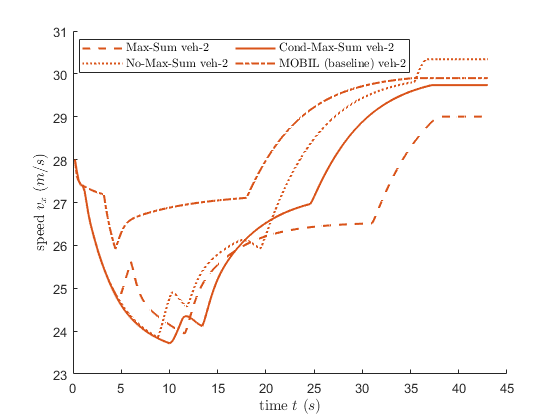}
    \caption{Speed trajectories for veh-2.}
    \label{fig:veh2}
    \Description{This figure displays the speed trajectories (over time) of agent veh-2. Each examined variant is displayed with different line style.}
\end{figure}

\begin{figure}
    \centering
    \includegraphics[width=0.45\textwidth]{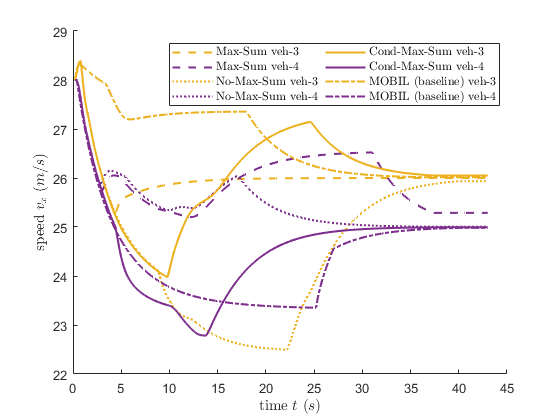}
    \caption{Speed trajectories for veh-3 and veh-4.}
    \label{fig:veh34}
    \Description{This figure displays the speed trajectories (over time) of agents veh-3 and veh-4. Each examined variant is displayed with different line style.}
\end{figure}

\begin{figure}
    \centering
    \includegraphics[width=0.45\textwidth]{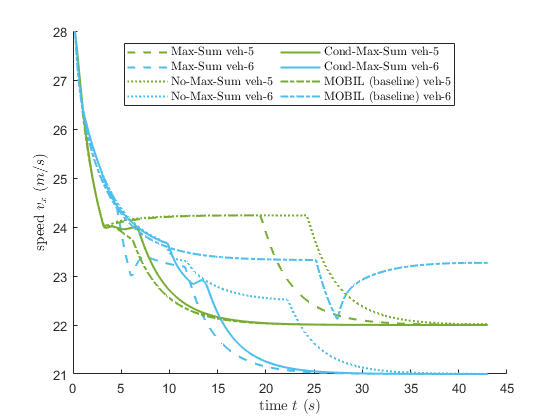}
    \caption{Speed trajectories for veh-5 and veh-6.}
    \label{fig:veh56}
    \Description{This figure displays the speed trajectories (over time) of agents veh-5 and veh-6. Each examined variant is displayed with different line style.}
\end{figure}

In addition, we further illustrate how each separate vehicle is affected by the different methods by showcasing all speed trajectories in Figs.~\ref{fig:veh2},~\ref{fig:veh34},~\ref{fig:veh56}.
In order to give higher priority for veh-1, this of course comes at a cost for the nearby vehicles due to the limited lateral space of the road.
This is more evident in veh-2's trajectories---veh-2 is closer to veh-1, both in terms of proximity and objective--- (Fig.~\ref{fig:veh2}), where Cond-Max-Sum appears to give it a lower priority in order to accommodate veh-1 (this can also be observed in the provided video). 
For the slowest vehicles veh-5,veh-6, we see that both Cond-Max-Sum and the standard Max-Sum algorithm provide similar results, with the remaining methods exhibiting worse behaviour.
Then, for vehicles veh-3,veh-4, which have a moderate desired speed objective, we see stronger fluctuations due to the fact that they need to both assist the faster vehicles veh-1,veh-2 in their overtake, and also overtake the slower vehicles veh-5,veh-6. 
There, we have a less clear distinction between all examined methods, e.g., Max-Sum is the best for veh-3, but at the same time the worst for veh-4, while the opposite is true for No-Max-Sum.
Cond-Max-Sum shows the biggest drop in speed for both vehicles, evidently due to the accommodation of veh-1,veh-2 that wish to overtake, while veh-3,veh-4 gave them space by staying behind the even slower vehicles veh-5,veh-6.
While this appears worse for these two individual vehicles,
it is more beneficial for the system (based on the average results) as a whole, given the way the factors are designed.

\begin{figure}
    \centering
    \includegraphics[width=0.45\textwidth]{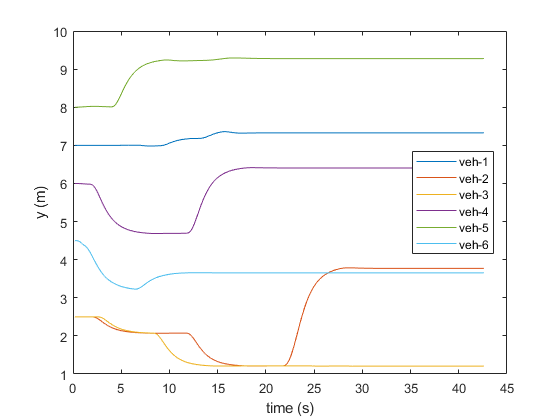}
    \caption{Lateral positioning of all vehicles with Cond-Max-Sum.}
    \label{fig:pos_y_cond}
    \Description{This figure displays the speed trajectories (over time) of agents veh-3 and veh-4. Each examined variant is displayed with different line style.}
\end{figure}

\begin{figure}
    \centering
    \includegraphics[width=0.45\textwidth]{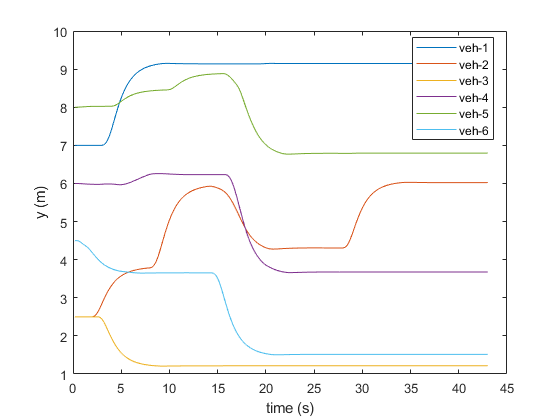}
    \caption{Lateral positioning of all vehicles with the standard Max-Sum algorithm.}
    \label{fig:pos_y_max}
    \Description{This figure displays the trajectories (over time) of all 6 agents regarding their lateral placement. Plot lines of each agent are associated based on the color choice. The trajectories stem from the application of the standard Max-Sum variant.}
\end{figure}

\begin{figure}
    \centering
    \includegraphics[width=0.45\textwidth]{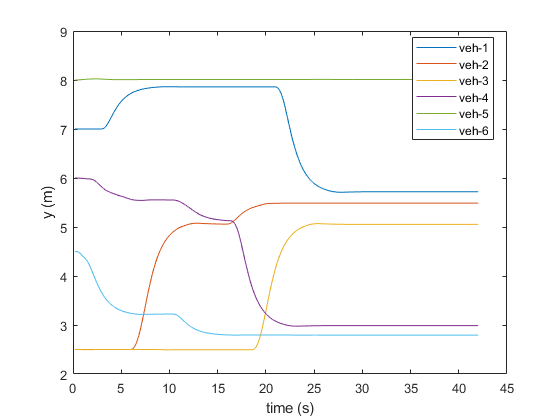}
    \caption{Lateral positioning of all vehicles with the No-Max-Sum algorithm.}
    \label{fig:pos_y_neutral}
    \Description{This figure displays the trajectories (over time) of all 6 agents regarding their lateral placement. Plot lines of each agent are associated based on the color choice. The trajectories stem from the application of the No-Max-Sum variant.}
\end{figure}

\begin{figure}
    \centering
    \includegraphics[width=0.45\textwidth]{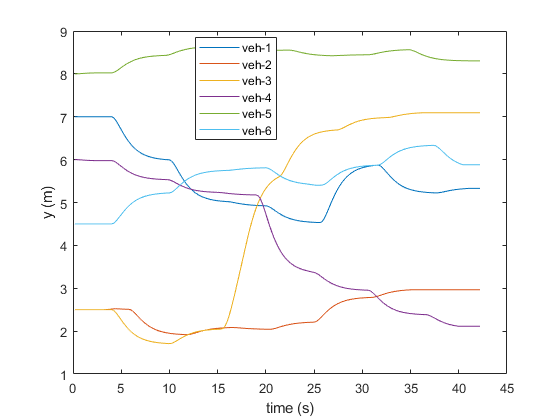}
    \caption{Lateral positioning of all vehicles with the MOBIL algorithm.}
    \label{fig:pos_y_mobil}
    \Description{This figure displays the trajectories (over time) of all 6 agents regarding their lateral placement. Plot lines of each agent are associated based on the color choice. The trajectories stem from the application of the MOBIL variant that does not rely on communication.}
\end{figure}

Moreover, in Figs.~\ref{fig:pos_y_cond},~\ref{fig:pos_y_max},~\ref{fig:pos_y_neutral},~\ref{fig:pos_y_mobil}, we show how the vehicles are positioned laterally for all examined methods. As also indicated by the values of the comfort-related metric of jerk in the main text, these further support the fact that Cond-Max-Sum achieves either similar or greater performance to Max-Sum but is also consistently combined with smoother lateral maneuvers due to the enhanced coordination among agents.

\begin{table}[ht]
  \centering
  
  \caption{Initial Timings for Lane-Free Coordination Problem}
  {
  \begin{tabular}{|c|c|}
  
  \hline
     Vehicle & Initial Time \\ \hline \hline
    veh-1 & $-1 sec$ \\ \hline
    veh-2 & $-2 sec$ \\ \hline
    veh-3 & $-1.4 sec$ \\ \hline
    veh-4 & $-2.4 sec$ \\ \hline
    veh-5 & $0 sec$ \\ \hline
    veh-6 & $-3.6 sec$ \\ \hline
  \end{tabular}
  }
  \label{tab:init_time}  
\end{table}

\subsection{Open and Large Lane-Free Environments}

The examined highway environment is a $2km$ length road with a $10.2m$ width, with $1h$ of simulation time.
Execution time of Cond-Max-Sum in a $15000 veh/h$ with approximately $300$ agents was measured on average around $41.59 msec$ (milliseconds) per time-step.
With the employed discrete step of $0.2 sec=200 msec$, we have $207.95msec$ per second of simulation time.
In the open environment, we have demand flow values at $10000, 15000 veh/h$, resulting in vehicles constantly entering the highway following that flow rate.
A snapshot of the simulation environment for the examined flow of $15000 veh/h$ can be found in Fig.~\ref{fig:highway_open}, where we see a top-view of the agents near the entry of the highway, and a new one entering at the entry point on the left.

\begin{figure*}
    \centering
    \includegraphics[width=.9\linewidth]{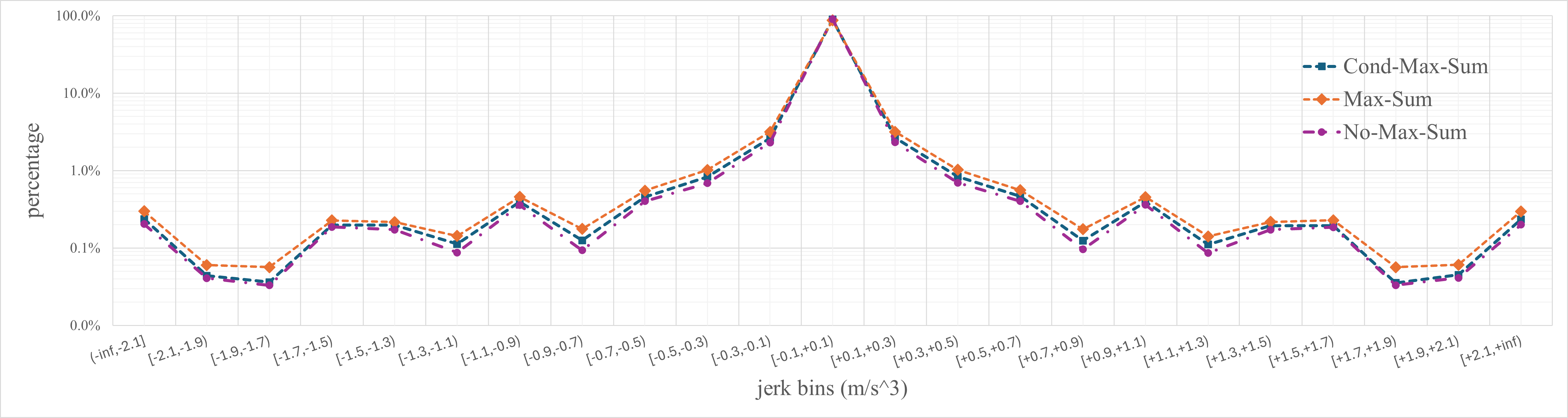}
    \caption{Lateral jerk percentage (histogram) plot for Cond-Max-Sum, Max-Sum and No-Max-Sum in $15000veh/h$.}
    \label{fig:hist_comfort}
    \Description{}
\end{figure*}

Finally, in Fig.~\ref{fig:hist_comfort}, we contain the full (histogram) plot of percentages for each jerk bin.
We observe consistent improvement of Cond-Max-Sum with respect to Max-Sum across all bins, since in the ideal case (jerk bin containing $0 m/s^3$), Cond-Max-Sum has higher percentage; whereas for all remaining bins indicating increasingly discomfort levels,  percentages are always in favor of Cond-Max-Sum.

\begin{table*}
  \centering
  
  \caption{Z-score and Confidence Intervals of Cond-Max-Sum and Max-Sum from speed measurements}
    \scalebox{0.75}
    {
    \begin{tabular}{lrrrrrrrrrrrr}
    \hline
     & $[0,5)$ & $[5,10)$ & $[10,15)$ & $[15,20)$ & $[20,25)$ & $[25,30)$ & $[30,35)$ & $[35,40)$ & $[40,45)$ & $[45,50)$ & $[50,55)$ & $[55,60]$ \\ 
    \midrule
    z-score & $\mathbf{2.56}$ & $\mathbf{4.01}$ & $\mathbf{3.32}$ & $\mathbf{4.82}$ & $\mathbf{8.03}$ & $\mathbf{10.30}$ & $\mathbf{6.46}$ & $\mathbf{4.60}$ & $\mathbf{8.16}$ & ${0.92}$ & $\mathbf{4.96}$ & ${1.24}$ \\
    CI (Cond-Max-Sum) & $2.92E-04$ & $1.67E-04$ & $1.53E-04$ & $1.57E-04$ & $1.88E-04$ & $1.46E-04$ & $1.70E-04$ & $2.39E-04$ & $1.82E-04$ & $2.59E-04$ & $1.21E-04$ & $1.35E-04$ \\
    CI (Max-Sum) & $2.95E-04$ & $1.43E-04$ & $1.59E-04$ & $1.82E-04$ & $2.06E-04$ & $1.54E-04$ & $1.87E-04$ & $2.28E-04$ & $1.84E-04$ & $2.44E-04$ & $1.35E-04$ & $1.46E-04$ \\
    \end{tabular}
    }
  \label{tab:stat}  
\end{table*}

\subsection{Statistical Significance}

For the speed measurements in Fig.~\ref{fig:speed_compare}, we have additionally performed a two-sample z-test (Cond-MS vs MS) for all the speed measurements in this figure.
Based on the calculated z-score, we verify that the difference is statistically significant with $a=0.05$ for $10$ out of the $12$ 5-minute intervals shown in the figure.
Confidence Interval (CI) is at $95\%$ (limit value $z=1.96$) for each $5$-minute period. 
The values are obtained from $n=1500$ samples during each period.
Standard Error (SE) information is not included, but can be inferred by CI,z,n values.
Z-scores and CI information are available in Table~\ref{tab:stat}.

\subsection{Measurements on Scalability and Communication Overhead}

In Table~\ref{tab:fg_scale}, we have measurements from the two scenarios of $10000$ and $15000$ $veh/h$.
Each agent is always associated with $1$ single factor; therefore, the corresponding size directly reflects the average number of agents, and we can also calculate the number of pairwise connections per agent (considering that $1$ pairwise factor connects $2$ agents/variables) as:

\begin{eqnarray}
    c_p = 2\cdot\frac{\#\text{Pairwise Factors}}{\#\text{Single Factors}}
\end{eqnarray}

Finally, we can calculate the size of messages, accounting for the variable space 
($|X_i|=15$, cf. Table~\ref{tab:fg}), that are exchanged between agents. In our approach, we believe it is fitting to measure the size of the broadcasted q messages accordingly and include the size for $x^*$ and time-estimates. 

As such, assuming a constant size of $N\, bytes$ per real value broadcasted, we can calculate the size of information $i_b$ in bytes to be broadcasted per agent:

\begin{eqnarray}
    i_b = (c_p \cdot |X_i| + 2)\cdot N\, bytes
\end{eqnarray}

\noindent with the first term $c_p \cdot |X_i|$ enumerating the total size of real values ($|X_i|$ values per message $q_{i\rightarrow j}$ to connected factor $j$), and the second containing the $2$ constant values ($x^*_i,t_{i,e}$) that are always shared.
As such, by restricting the number of connections, we can bound the amount of information given physical requirements. 

We also have two additional settings in situations with higher desired speed deviations (HDS) among agents (desired speed range is $[20,40] m/s=[72,144]km/h$ instead of $[25,35]m/s=[90,126]km/h$), to show how the graph’s size and pairwise connections are affected. We see that agents form more pairwise connections and thus connections on average, specifically for the flow of $15000$, we go from $\sim 7$ to $\sim 8$ connections per agent  due to more frequent overtakes. 
The higher speed deviations also affect the speed-related metrics negatively, hence the reason we observe more agents (single factors) in the road when comparing the two cases at the same flow.
Measurements below are taken from Cond-Max-Sum, with similar results for all variants in that regard.

\begin{table}[t]
  \centering
  
  \caption{Scalability Measurements on the Factor Graph}
  {
  \begin{tabular}{lrr}
  
  \hline
     Scenario & \#Single Factors (avg) &	\#Pairwise Factors (avg) \\ 
    \midrule
    $10000 veh/h$ &	$187.35$ &	$445.04$ \\
    $15000 veh/h$ &	$287.14$ &	$1017.66$ \\
    $10000 veh/h$ (HDS) &	$202.16$ &	$575.69$ \\
    $15000veh/h$ (HDS) &	$315.66$ &	$1263.80$ \\
    
  \end{tabular}
  }
  \label{tab:fg_scale}  
\end{table}
\end{document}